\documentclass[12pt,a4paper]{scrartcl}

\usepackage[utf8]{inputenc}
\usepackage[T1]{fontenc}
\usepackage{amsmath,amssymb,mathtools}
\usepackage[separate-uncertainty=true,retain-unity-mantissa=false]{siunitx}
\usepackage{slashed}
\usepackage{graphicx}
\usepackage{booktabs,multicol,multirow}
\usepackage[shortlabels]{enumitem}
\usepackage{hyperref}
\hypersetup{
  colorlinks   = true, 
  urlcolor     = blue, 
  linkcolor    = black,
  citecolor   = blue 
}

\usepackage[capitalize]{cleveref}
\usepackage{xspace}
\usepackage[noblocks]{authblk}
\usepackage{accents}
\usepackage[dvipsnames]{xcolor}
\usepackage{soul}
\usepackage[textsize=tiny]{todonotes}
\usepackage{subcaption}
\usepackage[normalem]{ulem}
\usepackage{bbold}
\usepackage[sorting=none,%
  citestyle=numeric-comp,%
  bibstyle=mynumeric,%
  backend=bibtex,%
  giveninits=true]{biblatex}
\newcommand{\fb}{\ensuremath{\mathrm{fb}}\xspace}
\newcommand{\pb}{\ensuremath{\mathrm{pb}}\xspace}
\newcommand{\GeV}{\ensuremath{\mathrm{GeV}}\xspace}
\newcommand{\TeV}{\ensuremath{\mathrm{TeV}}\xspace}
\newcommand{\MeV}{\ensuremath{\mathrm{MeV}}\xspace}
\newcommand{\nn}{\nonumber}

\newcommand{\mmiss}{\ensuremath M_{\text{miss}}}
\newcommand{\etmiss}{\ensuremath E^T_\text{miss}}

\newcommand{\lb}{\left(}
\newcommand{\rb}{\right)}

\newcommand{\lam}{\lambda}

\NewBibliographyString{refname}
\NewBibliographyString{refsname}
\DefineBibliographyStrings{english}{%
  refname = {Ref\adddot},
  refsname = {Refs\adddot}
}
\DeclareCiteCommand{\ccite}
  {%
  \ifnum\thecitetotal=1
    \bibstring{refname}%
  \else%
    \bibstring{refsname}%
  \fi%
  \addnbspace\bibopenbracket%
  \usebibmacro{cite:init}%
   \usebibmacro{prenote}}
  {\usebibmacro{citeindex}%
   \usebibmacro{cite:comp}}
  {}
  {\usebibmacro{cite:dump}%
   \usebibmacro{postnote}%
   \bibclosebracket}
\newrobustcmd*{\Ccite}{\bibsentence\ccite}

\newcommand{\madgraph}{{\textsc{MadGraph5\textunderscore aMC@NLO}}}
\newcommand{\sklearn}{{\textsc{scikit-learn}}}

\newcommand{\xgboost}{{\textsc{XGBoost}}}

\newcommand*{\pbar}[1]{\accentset{\scalebox{0.4}{$(-)$}}{#1}}

\newcommand{\edit}[1]{{\color{black}#1}}

\addtokomafont{disposition}{\rmfamily}

\title{\Large 
Probing the Inert Doublet Model via Vector-Boson Fusion at a Muon Collider
}
\author[1]{Johannes Braathen\footnote{\href{mailto:johannes.braathen@desy.de}{johannes.braathen@desy.de}}}
\author[1,2]{Martin Gabelmann\footnote{\href{mailto:martin.gabelmann@desy.de}{martin.gabelmann@physik.uni-freiburg.de}}}
\author[3]{Tania Robens\footnote{\href{mailto:trobens@irb.hr}{trobens@irb.hr}}}
\author[1]{Panagiotis Stylianou\footnote{\href{mailto:panagiotis.stylianou@desy.de}{panagiotis.stylianou@desy.de}}}
\affil[1]{Deutsches Elektronen-Synchrotron DESY, Notkestr.~85, 22607 Hamburg, Germany}
\affil[2]{Albert-Ludwigs-Universität Freiburg, Physikalisches Institut, Hermann-Herder-Str. 3, 79104 Freiburg, Germany}
\affil[3]{Ruder Boskovic Institute, Bijenicka cesta 54, 10000 Zagreb, Croatia}

\vspace{-10mm}
\date{{\small\today}}
\titlehead{\hfill \parbox{5cm}{\vspace{-2cm}\begin{flushright} \texttt{DESY-24-170}\\
\texttt{FR-PHENO-2024-009}\\
\texttt{RBI-ThPhys-2024-25}\end{flushright}}}

\begin{document}
\maketitle

\begin{abstract}\noindent
\vspace{-12mm}\\
    In this work, we explore the discovery potential of the Inert Doublet Model (IDM) via the vector boson fusion (VBF) channel at a muon collider with centre-of-mass energy of 10 \TeV. The Inert Doublet Model is a two-Higgs-doublet model variant with an unbroken discrete $\mathbb{Z}_2$ symmetry, featuring new stable scalar particles that can serve as dark matter candidates. Current dark matter data constrain the phenomenologically viable parameter space of the IDM and render certain collider signatures elusive due to tiny couplings. However, VBF-type processes can still exhibit significant enhancements compared to the Standard Model, presenting a promising avenue to probe the IDM at a high-energy muon collider. We consider as our specific target process $\mu^+\mu^-\to \nu_\mu\bar{\nu}_\mu AA\to \nu_\mu\bar{\nu}_\mu jj \ell\ell HH$, where $H$ and $A$ are the lightest and second-lightest new scalars and $\ell$ can be electrons or muons. We perform both cut-based and machine-learning improved sensitivity analyses for such a signal, finding a population of promising benchmark scenarios. We additionally investigate the impact of the collider energy by comparing sensitivities to the target process at 3 \TeV and 10 \TeV. Our results provide a clear motivation for a muon collider design capable of reaching a 10 \TeV centre-of-mass energy. We furthermore discuss  constraints stemming from new-physics corrections to the Higgs to di-photon decay rate as well as the trilinear Higgs coupling in detail, using state-of-the-art higher-order calculations.
\end{abstract}
\setcounter{footnote}{0}

\newpage

\tableofcontents
\hypersetup{linkcolor=blue}

\newpage


\section{Introduction}
\label{sec:intro}

The current collider landscape promises quite a few options for the discovery --- or exclusion --- of new physics scenarios in various realisations. However, some of these scenarios remain elusive at hadron machines due to weak couplings implying small rates, and are not directly accessible at low-energy lepton colliders as they feature new physics states with relatively large masses. For such scenarios, one might instead turn to higher-energy lepton colliders, that provide both access to heavier mass states as well as a relatively clean collider environment.

In this work, we investigate the production of new physics states at muon colliders. The idea of a muon collider has recently experienced an incredible revival, documented in various community efforts, including the recent Snowmass activities (see e.g. Refs.~\cite{Black:2022cth,Aime:2022flm,Accettura:2023ked,InternationalMuonCollider:2024jyv,MuCoL:2024oxj}). A particular feature of the new generation of proposed muon collider facilities is that they can in principle go up to relatively larger centre-of-mass energies on the multi-TeV scale, while at the same time featuring a  cleaner environment than hadron colliders. At such high energies, electroweak gauge bosons can effectively be considered massless, leading to a logarithmic enhancement of colinear and soft emissions of such particles. In this way, vector-boson fusion (VBF) processes can in principle be largely enhanced, as discussed for example in Refs.~\cite{Dawson:1984gx,Kane:1984bb,Kunszt:1987tk,Chanowitz:1984ne,Gunion:1986gm}, and more recently in Refs.~\cite{Chen:2016wkt,Ruiz:2021tdt,Costantini:2020stv,Han:2020uid,Garosi:2023bvq,Denner:2024yut}\footnote{See also \cite{Buttazzo:2018qqp} for related work.}.

We here concentrate on the investigation of the Inert Doublet Model (IDM) \cite{Deshpande:1977rw,Barbieri:2006dq,Cao:2007rm}, a two scalar doublet model that obeys an unbroken discrete $\mathbb{Z}_2$ symmetry. One of the two scalar $SU(2)_L\times\,U(1)_Y$ doublets acts like the doublet of the Standard Model (SM) and is responsible for the electroweak symmetry breaking (EWSB). The lightest scalar of the second doublet is the dark matter (DM) candidate which, at least in some regions of the parameter space, can provide the full relic density as measured by the PLANCK experiment \cite{Planck:2018vyg}. The model has been vastly discussed in the literature, see e.g. Refs.~\cite{Swiezewska:2012eh,Gustafsson:2012aj,Arhrib:2013ela,Krawczyk:2013jta,Ilnicka:2015jba,Diaz:2015pyv,Belyaev:2016lok,Eiteneuer:2017hoh,Ilnicka:2018def,Kalinowski:2018ylg,Banerjee:2019luv,Tsai:2019eqi,Fabian:2020hny,Kalinowski:2020rmb,Basu:2020qoe,Banerjee:2021xdp,Banerjee:2021hal,Banerjee:2021anv,AbdusSalam:2022idz,Astros:2023gda,Justino:2024etz} for work that includes general scans and updated constraints, as well as Refs.~\cite{Poulose:2016lvz,Kanemura:2016sos,Datta:2016nfz,Wan:2018eaz,Belyaev:2018ext,Heisig:2018kfq,Kalinowski:2018kdn,Dercks:2018wch,Bhardwaj:2019mts,Guo-He:2020nok,Abouabid:2020eik,Yang:2021hcu,Ghosh:2021noq,Klamka:2022ukx,Fan:2022dck,Abouabid:2022rnd,He:2024bwh,Phan:2024vfy} for dedicated collider studies and recasts taking the 125-\GeV~Higgs mass into account. 

As a concrete target, in this work we study the VBF-type pair production of the second neutral scalar boson, $A$, in association with missing energy stemming from both neutrinos as well as the dark matter candidates $H$ that the $A$ scalars decay into. The production of these particles is largely suppressed at hadron machines due to direct detection constraints that severely limit the respective new physics coupling, see e.g. Ref.~\cite{Kalinowski:2020rmb} for a comparison of possible collider rates for various production modes within the Inert Doublet Model. However, at high-energy lepton colliders, the respective VBF-type production process is largely enhanced. In this work, we find a variety of benchmark scenarios of the IDM that can be probed with the VBF-type $A$ pair production process at a 10 \TeV muon collider --- reaching significances of 5 or more employing a Machine-Learning (ML) improved analysis. 

This paper is organised as follows: in~\cref{sec:model} we briefly introduce the IDM and then discuss the different theoretical and experimental constraints that we take into consideration throughout this work. We subsequently provide details on our parameter scan and the different benchmark points (BPs) in~\cref{seq:procgen} while also motivating the signal process under investigation. Our analysis strategy is described in~\cref{sec:analysis} where we also comment on the difference between $10$ and $3$~TeV centre-of-mass energies, before concluding in~\cref{sec:conclusions}. 

\section{Model}
\label{sec:model}

\subsection{Definitions and conventions}
The IDM~\cite{Deshpande:1977rw,Barbieri:2006dq,Cao:2007rm} adds a second $SU(2)_L$ doublet of hypercharge $Y=1/2$ to the particle content of the SM. This additional doublet is charged under an unbroken $\mathbb{Z}_2$ symmetry, while all other, SM-like, states transform trivially under this symmetry. The IDM differs from standard realisations of Two-Higgs-Doublet Models because the $\mathbb{Z}_2$ symmetry remains unbroken, even after electroweak symmetry breaking, which also implies that the new beyond-the-SM (BSM) scalars do not mix with SM-like states, nor do they couple to fermions --- for these reasons, the BSM scalars are usually referred to as \textit{inert scalars} in the IDM. 

The SM-like and the new doublets, denoted respectively $\Phi_1$ and $\Phi_2$, can be expanded as
\begin{align}
    \Phi_1=\begin{pmatrix}
      G^+\\
      \frac{1}{\sqrt{2}}(v+h+iG^0)
    \end{pmatrix}\,,
    \quad \text{and} \quad
    \Phi_2=\begin{pmatrix}
      H^+\\
      \frac{1}{\sqrt{2}}(H+iA)
    \end{pmatrix}\,,
\end{align}
where $h$ denotes the discovered Higgs boson with a mass of 125 GeV, $G^0$ and $G^\pm$ are the neutral and charged would-be Goldstone bosons, and $H$, $A$, and $H^\pm$ are the inert scalars. The lightest of the inert scalars is stable; it therefore constitutes a candidate for dark matter and is treated as invisible in collider processes. Throughout this paper, we will  consider $H$ to be the lightest scalar.\footnote{In principle, one can also choose the charged inert scalar to be the DM candidate. However, there are strong constraints on such scenarios \cite{ParticleDataGroup:2024cfk}, and we therefore disregard this possibility in this work.}  

The tree-level scalar potential of the theory can be written as 
\begin{align}
    V^{(0)}_\text{IDM} =&\ \mu_1^2 \big|\Phi_1\big|^2 + \mu_2^2 \big|\Phi_2\big|^2 + \frac{1}{2} \lambda_1 \big|\Phi_1\big|^4 + \frac{1}{2} \lambda_2 \big|\Phi_2\big|^4 + \lambda_3 \big|\Phi_1\big|^2 \big|\Phi_2\big|^2 + \lambda_4 \big|\Phi_1^\dagger \Phi_2\big|^2 \nn\\
    &+ \frac{1}{2} \lambda_5 \left[\big(\Phi_1^\dagger\Phi_2\big)^2+\text{h.c.}\right]\,. 
\end{align}
All parameters in this potential can be taken to be real. In particular, a phase of $\lambda_5$ can always be rotated away by an $SU(2)_L$ transformation of $\Phi_2$, and our choice of $H$ as DM candidate corresponds to taking $\lambda_5$ negative (we note that the roles of $H$ and $A$ can be in principle exchanged by flipping the sign of $\lambda_5$). 

After EWSB, the mass parameter $\mu_1$ can be eliminated using the tadpole equation $\mu_1^2+\frac{1}{2}\lambda_1v^2=0$. On the other hand, $\mu_2$ is the mass parameter that controls the decoupling of the inert scalars. 
The tree-level scalar masses read 
\begin{align}
\label{eq:IDM_0Lmasses}
    M_h^2 &= \lambda_1 v^2, \,\nn\\
    M_H^2 &= \mu_2^2 + \frac{1}{2}\lambda_{345} v^2,\,\nn\\
    M_A^2 &= \mu_2^2 + \frac{1}{2}\bar{\lambda}_{345}v^2,\,\nn\\
    M_{H^\pm}^2 &= \mu_2^2 + \frac{1}{2}\lambda_3 v^2,\,
\end{align}
where we have defined the short-hand notations $\lambda_{345}\equiv \lambda_3+\lambda_4+\lambda_5$ and $\bar{\lambda}_{345}\equiv \lambda_3+\lambda_4-\lambda_5$. 

In addition to $v$ and $M_h$, the scalar sector of the IDM can be described in terms of five free parameters, which we choose to be
\begin{align}
    M_H,\ M_A,\ M_{H^\pm},\ \lambda_2,\ \text{and} \ \lambda_{345}\,. 
\end{align}
We note that the other commonly-used parameter $\bar{\lambda}_{345}$ can be obtained from the above set of parameters via the relation
\begin{align}
\label{eq:l345b}
\bar\lambda_{345}=\lambda_{345} +\frac{2(M_A^2-M_H^2)}{v^2}\,.
\end{align}

\subsection{Theoretical and experimental constraints}
\label{sec:thexpconstraints}
The allowed parameter space of the IDM is subject to various theoretical and experimental constraints. We review them briefly in this section. Most of the constraints follow the implementation and prescriptions as given in \cite{Ilnicka:2015jba,Kalinowski:2020rmb}.

\paragraph*{Inert vacuum condition:} In order for the vacuum in which $\Phi_2$ does not acquire a vacuum expectation value to be a global minimum of the potential, and thus for the $\mathbb{Z}_2$ symmetry to remain unbroken after EWSB, the following condition~\cite{Ginzburg:2010wa} must be fulfilled at leading order (LO)
\begin{align}
    \frac{\mu_2^2}{\sqrt{\lambda_2}}\geq \frac{\mu_1^2}{\sqrt{\lambda_1}}\,.
\end{align}

\paragraph*{Boundedness-from-below of the potential:} A second condition is that the scalar potential should remain bounded from below. 
One can show that this leads to the conditions on the quartic couplings~\cite{Deshpande:1977rw,Kanemura:1999xf}
\begin{align}
    \lambda_1 \geq 0,\quad \lambda_2\geq 0,\quad \sqrt{\lambda_1\lambda_2}+\lambda_3+\mathrm{min}\big\{0,\lambda_4\pm\lambda_5\big\}\geq 0\,.
\end{align}

\paragraph*{Perturbative unitarity:} Another theoretical constraint that we consider is perturbative unitarity. We employ in our work both tree-level results from Refs.~\cite{Kanemura:1993hm,Akeroyd:2000wc}, via the public code \texttt{2HDMC}~\cite{Eriksson:2009ws}, as well as one-loop results from Refs.~\cite{Cacchio:2016qyh,Grinstein:2015rtl}.

\paragraph*{Perturbativity of the couplings:} 
Finally, we check for perturbativity of the couplings. For this we require all quartic scalar couplings (in the mass basis) to have absolute values lower than $4\,\pi$, and require the same for the couplings in the potential. The former check is again performed employing \texttt{2HDMC}.

\paragraph*{Gauge boson widths and electroweak precision observables:} Turning now to constraints arising from experimental results, we must ensure that the decay widths of the $W$ and $Z$ gauge bosons, which are measured very precisely~\cite{ParticleDataGroup:2022pth}, are not drastically modified by the opening of new BSM decay channels, such as $W^\pm\to HH^\pm$ or $AH^\pm$, or $Z\to HA$ or $H^+H^-$. Requiring that these new channels remain kinematically forbidden leads to the inequalities
\begin{align}
    M_H+M_{H^\pm}\geq M_W\,,\quad M_A+M_{H^\pm}\geq M_W\,,\quad M_H+M_A\geq M_Z\,,\quad 2M_{H^\pm}\geq M_Z\,.
\end{align}

Moreover, electroweak precision observables (EWPO) provide stringent constraints on the IDM parameter space, in particular in terms of allowed mass splittings. In this work, we follow the common choice of parametrising the EWPO via the oblique parameters $S,\ T,\ U$~\cite{Altarelli:1990zd,Peskin:1990zt,Peskin:1991sw,Maksymyk:1993zm}, for which fit results have been obtained by the \texttt{GFitter} collaboration~\cite{Haller:2018nnx}. All benchmark points considered in the following have been required to fulfill a $2\sigma$ level of agreement in the oblique parameters with the allowed ranges from these fit results. The necessary calculations of $S,\ T,\ U$ were performed with the public tool \texttt{2HDMC}.

\paragraph*{Constraints from dark matter:} We also take into account the DM relic density as well as results from direct detection experiments in the checks of our benchmark scenarios. The relic density $\Omega_H h^2$ and direct-detection cross-section $\sigma_{DD}(M_H)$ of the DM candidate $H$ are computed\footnote{A comparison of results using different versions and settings of \texttt{micrOMEGAs} for a set of benchmark points can be found in \cite{Kalinowski:2020rmb}. } with the public tool \texttt{micrOMEGAs\_5.0.4}~\cite{Belanger:2018ccd}. In order not to overclose the Universe, it is required to fulfill the inequality 
\begin{align}
    \Omega_H h^2\leq \Omega_c h^2 = 0.1200 \pm 0.0012
\end{align}
where $\Omega_c h^2$ is the DM relic density determined with PLANCK data~\cite{Planck:2018vyg}. We take this condition as an inequality, rather than an equality, meaning that we allow the possibility that $H$ is not the unique component constituting DM. Scenarios where the IDM can provide the exact relic density have e.g. been discussed in Ref.~\cite{Kalinowski:2020rmb}. 

Following Refs.~\cite{Ilnicka:2015jba,Kalinowski:2018ylg}, predictions for the direct detection cross-section are tested against limits from the LUX-ZEPLIN experiment~
\cite{LZCollaboration:2024lux}, taking into account a rescaling of the cross-section for multi-component DM scenarios ($i.e.$ when $\Omega_Hh^2< \Omega_ch^2$). In practice, we verify that
\begin{align}
    \sigma_{DD}(M_H)\,\frac{\Omega_H}{\Omega_c}\leq \sigma^\text{LZ}_\text{lim} (M_H)\,.
\end{align}

\paragraph*{Collider searches at LEP and LHC:}
Collider searches have so far only produced null results for BSM scalars, which we take into account for our benchmark scenarios. Reinterpretations of LEP searches for supersymmetric particles provide lower bounds on the masses of the inert scalars: specifically, searches reinterpreted for the $e^+e^-\to H^+H^-$ process~\cite{Pierce:2007ut} yield
\begin{align}
    M_{H^\pm}\gtrsim 70\text{ GeV}\,,
\end{align}
while a recast of SUSY searches for neutralino pair-production to the $e^+e^-\to H A$ process~\cite{Lundstrom:2008ai}, which would produce a visible di-jet or di-lepton final state, excludes the region of the IDM parameter space where simultaneously
\begin{align}
    M_H \leq 80\text{ GeV} \text{ and } M_A\leq 100\text{ GeV} \text{ and } |M_H-M_A| \geq 8\text{ GeV}\,. 
\end{align}

Recasts of different LHC searches for mono-jet signals~\cite{Belyaev:2018ext}, for signals with two~\cite{Belanger:2015kga,Belyaev:2022wrn} or more~\cite{Belyaev:2022wrn} leptons plus missing transverse energy and for invisible Higgs decays in VBF production~\cite{Dercks:2018wch} provide constraints on the IDM parameter space. It should however be noted that the regions where recasts of LHC data offer sensitivity are typically already in tension with dark matter data (in particular the DM relic density). We are also checking against \texttt{HiggsBounds} \cite{Bechtle:2008jh,Bechtle:2011sb,Bechtle:2013wla,Bechtle:2020pkv} and \texttt{HiggsSignals} \cite{Bechtle:2013xfa,Bechtle:2014ewa,Bechtle:2020uwn}. Note that only the latter can give constraints on the model due to the fact that the current versions of \texttt{HiggsBounds} do not include final states with dark matter candidates that would require dedicated two-dimensional limit grids, as required for our studies\footnote{We thank T. Biek\"otter for useful discussions regarding this.}. Furthermore, the main quantities that influence the signal strength are the decays of Higgs to invisible as well as modifications of the di-photon rate. We will comment on these separately below. 

\begin{figure}
    \centering
    \includegraphics[width=\textwidth]{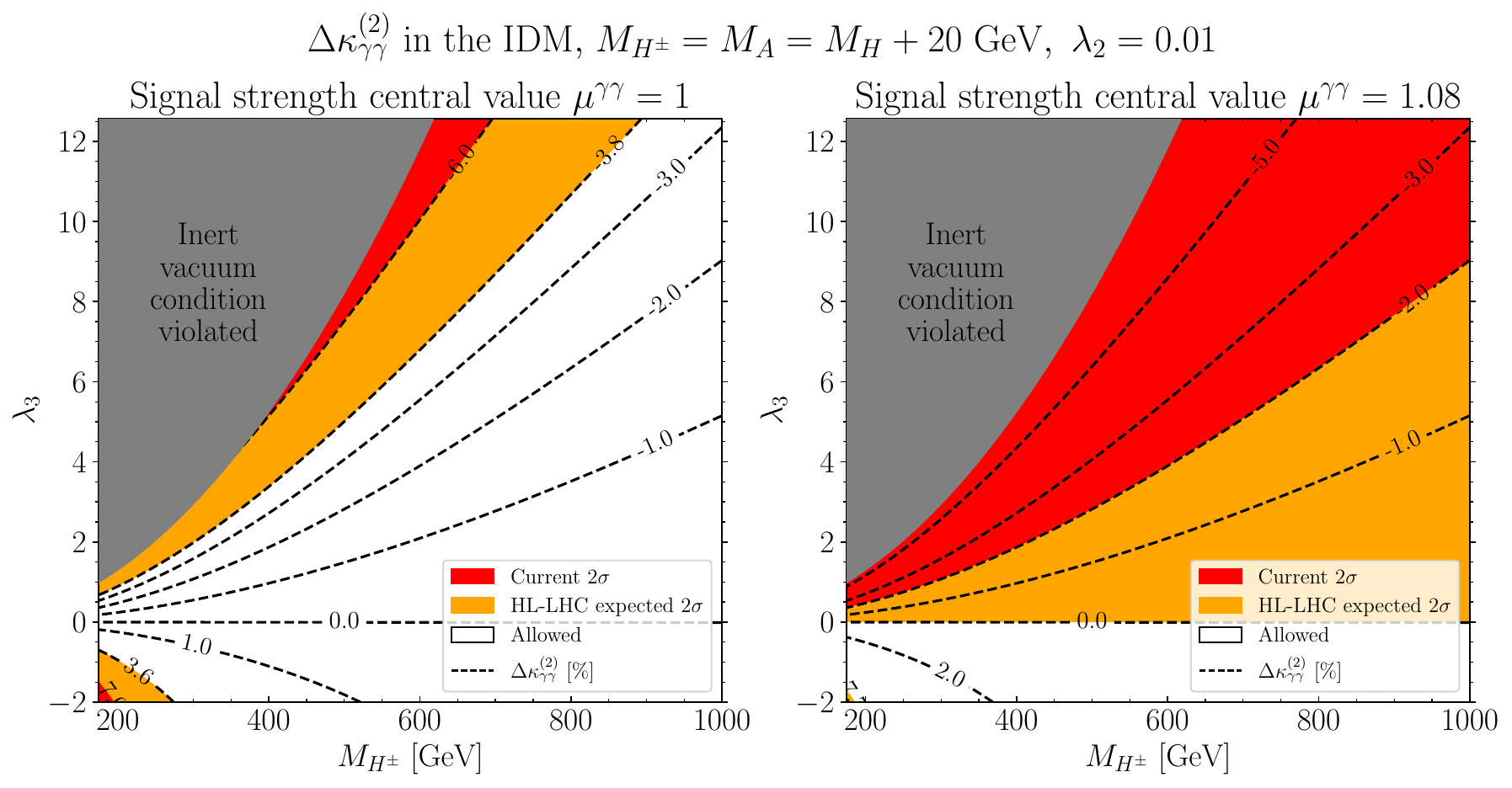}
    \caption{Contour lines of the BSM deviation in the Higgs decay width to two photons, shown in \%, and computed at two loops using expressions from Ref.~\cite{Aiko:2023nqj}, in the plane of the charged Higgs mass $M_{H^\pm}$ and the coupling $\lambda_3$. The differently shaded regions correspond to points that are already excluded at the $2\sigma$ level by current LHC results (red), points that are expected to be probed at the HL-LHC (orange), and points that will remain allowed even after HL-LHC (white), respectively. The left and right figures differ only by the central value assumed for the $\gamma\gamma$ signal strength: the left plot assumed a central value of $\mu^{\gamma\gamma}=1$ while the right plot takes the current experimental average of $\mu^{\gamma\gamma}=1.08$, and the corresponding $2\sigma$ bounds are discussed in the text. We impose a mass difference of 20 GeV between $M_{H^\pm}=M_A$ and $M_H$ (the mass of the DM candidate) and set $\lambda_2=0.01$. }
    \label{fig:Deltakappagamgam}
\end{figure}

\paragraph*{Properties of the 125-GeV Higgs boson:} 
A first property of the 125-GeV Higgs boson that provides a strong constraint on the IDM parameter space is its decay width to two photons --- see $e.g.$ Refs.~\cite{Kanemura:2016sos,Arhrib:2012ia,Swiezewska:2012eh,Aiko:2023nqj} for studies of this decay in the IDM. This decay is sensitive to effects from BSM scalars, and in particular the charged inert scalar, already at leading order ($i.e.$ one loop). Typical BSM deviations in the corresponding effective coupling, defined as
\begin{align}
\label{eq:def_Deltakapgamgam}
    \Delta\kappa_{\gamma\gamma}\equiv \sqrt{\frac{\Gamma(h\to\gamma\gamma)_\text{IDM}}{\Gamma(h\to\gamma\gamma)_\text{SM}}}-1\,,
\end{align}
are of the order of a few percent, which is similar to the current precision with which the coupling is constrained by LHC data, as well as future prospects at the HL-LHC. This is illustrated in  \cref{fig:Deltakappagamgam}, which shows the BSM deviation in the effective Higgs-photon-photon coupling $\Delta\kappa_{\gamma\gamma}$, determined from a calculation up to dominant two-loop level~\cite{Aiko:2023nqj}, in the plane of $(M_{H^\pm},\lambda_3)$. For illustration purposes, we here fixed $M_{H^\pm}=M_A$ and $M_H=M_{H^\pm}-20\text{ GeV}$, and also set $\lambda_2\,=\,0.01$. We note that we choose $M_{H^\pm}$ and $\lambda_3$ as axes of the parameter plane because these are the parameters that enter the prediction for $\Delta\kappa_{\gamma\gamma}$ from the leading (one-loop) order; the coupling $\lambda_{345}$ can straightforwardly be derived using \cref{eq:IDM_0Lmasses}. 
The left and right plot of \cref{fig:Deltakappagamgam} present the current and expected future bounds on $\Delta\kappa_{\gamma\gamma}$ for two different central values of the signal strength: the left plot assumes a SM-like value of $\mu^{\gamma\gamma}=1$, while the right plot uses the current experimental average of $\mu^{\gamma\gamma}=1.08$. The latter number is obtained by a naive combination of the signal strengths reported for the Higgs di-photon decay channel by ATLAS~\cite{ATLAS:2022tnm}, $\mu_\text{ATLAS}^{\gamma\gamma}=1.04^{+0.1}_{-0.09}$, and by CMS~\cite{CMS:2021kom}, $\mu_\text{CMS}^{\gamma\gamma}=1.12\pm0.09$, which yields $\mu_\text{comb.}^{\gamma\gamma}=1.08^{+0.07}_{-0.06}$, leading to $\Delta \kappa_{\gamma \gamma}\,\in\left[-2\%;10\%\right]$ at 2 standard deviations. The asymmetry in bounds stems from the current enhanced central value. For future limits, we take the expected $1\sigma$ bounds on $\mu^{\gamma\gamma}$ of $6\%$ for ATLAS and $4.4\%$ for CMS from Ref.~\cite{Cepeda:2019klc}. A naive combination would then lead to $\mu\,=\,1\,\pm\,0.037$, if we assume a SM like central value, leading\footnote{If the central value $\mu^{\gamma\gamma}$ remains at its current value, we instead would have $\left[0\%; 7.7\%\right]$ as the allowed interval.} to $\Delta\,\kappa_{\gamma\gamma}\,\in\,\left[-3.8\%;3.6\%\right]$ at the $2\sigma$ level. All numbers for the di-photon signal strength have been converted into numbers for $\Delta\kappa_{\gamma\gamma}$ according to \cref{eq:def_Deltakapgamgam}. 

The red shaded regions in \cref{fig:Deltakappagamgam} are outside the current $2\sigma$ experimental uncertainty band, while the orange shaded regions are outside the expected $2\sigma$ band at the HL-LHC. 
We note that while $M_{H^\pm}$ and $\lambda_3$ are the only free IDM parameters determining $\Delta\kappa_{\gamma\gamma}$ at one loop ($i.e.$ LO), at next-to-leading order (NLO), $M_H$, $M_A$ and $\lambda_2$ also enter the prediction for $\Delta\kappa_{\gamma\gamma}$. 

\begin{figure}
    \centering
    \includegraphics[width=.7\textwidth]{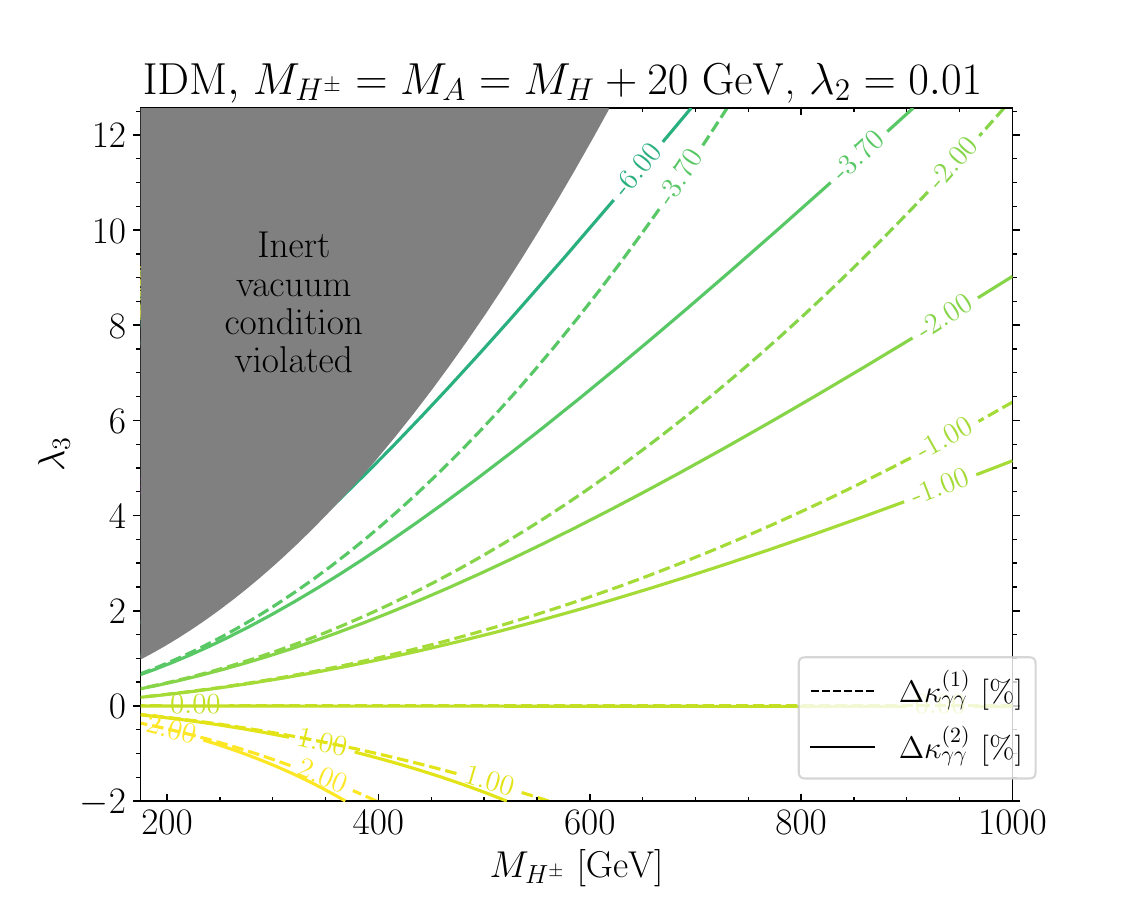}
    \caption{Comparison of contour lines for $\Delta\kappa_{\gamma\gamma}$ (shown in \%), computed in the IDM at one loop (dashed lines) and two loops (solid lines), in the $(M_{H^\pm},\lambda_3)$ plane{, for the same scenario as in \cref{fig:Deltakappagamgam}}. }
    \label{fig:Deltakappagamgam_1Lvs2L}
\end{figure}

To illustrate the relevance of including dominant two-loop corrections in $\Delta\kappa_{\gamma\gamma}$, we present in \cref{fig:Deltakappagamgam_1Lvs2L} contours for this quantity (in \%) computed at both one loop (dashed lines) and two loops (solid lines), for the same type of scenario as in \cref{fig:Deltakappagamgam}. For a given charged Higgs mass $M_{H^\pm}$, we observe that the bound from $\Delta\kappa_{\gamma\gamma}$ constrains the parameter space to lower values of the coupling $\lambda_3$ at two loops, which can be understood from the fact that both one- and two-loop BSM corrections contribute to the di-photon signal with the same, negative, sign. For smaller coupling values, the difference between one- and two-loop contours becomes smaller, as expected. 

A second property of the 125-GeV Higgs boson that offers a powerful probe~\cite{Bahl:2022jnx} of the IDM parameter space is its self-coupling $\lambda_{hhh}$. The current experimental bounds on this coupling, {obtained mostly from searches for di-Higgs production (with additional information from single-Higgs production measurements)}, are significantly looser than the ones on the di-photon decay, or other Higgs couplings. Upper limits on the di-Higgs cross-section can be translated into bounds on the coupling modifier $\kappa_\lambda$, defined as \begin{align}
    \kappa_\lambda \equiv \frac{\lambda_{hhh}}{(\lambda_{hhh}^\text{SM})^{(0)}}\,,
\end{align} 
where $(\lambda_{hhh}^\text{SM})^{(0)}\,\equiv\,3\,m_h^2/v$ is the tree-level prediction for the trilinear Higgs coupling in the SM.\footnote{ We use the normalisation $\mathcal{L}\supset -\frac{1}{6}\lambda_{hhh}h^3$ of the trilinear Higgs coupling throughout this paper.} The respective ATLAS and CMS observed bounds at $95\%$ Confidence Level (CL) are given by $\kappa_\lam\,\in\left[ -1.4; 6.1\right]$ \cite{ATLAS:2022jtk} and $\kappa_\lam\,\in\left[ -1.4; 7.8\right]$ \cite{CMS:2024awa} for a combination of di-Higgs and single-Higgs final states 
{with} all other relevant coupling modifiers 
floating, and  $\kappa_\lam\,\in\left[ -1.2; 7.2\right]$ \cite{ATLAS:2024ish} from di-Higgs searches only.
Meanwhile, it is known~\cite{Kanemura:2002vm,Kanemura:2004mg,Aoki:2012jj,Kanemura:2015fra,Kanemura:2015mxa,Arhrib:2015hoa,Kanemura:2016sos,Kanemura:2016lkz,He:2016sqr,Kanemura:2017wtm,Kanemura:2017wtm,Kanemura:2017gbi,Chiang:2018xpl,Basler:2018cwe,Senaha:2018xek,Braathen:2019pxr,Braathen:2019zoh,Kanemura:2019slf,Basler:2020nrq,Braathen:2020vwo,Bahl:2022jnx,Bahl:2022gqg,Falaki:2023tyd,Bahl:2023eau,Aiko:2023xui,Basler:2024aaf,Heinemeyer:2024hxa} that this coupling can exhibit significant BSM deviations, of several tens or even hundreds of percent, for points that are allowed by all relevant state-of-the-art theoretical and experimental constraints. As shown in Refs.~\cite{Bahl:2022jnx,Bahl:2023eau}, the current bounds on $\kappa_\lambda$ are already sufficiently stringent to constrain otherwise unconstrained regions of the parameter space of various BSM models, including the IDM. Moreover, the experimental limits on $\kappa_\lambda$ will be considerably improved at the HL-LHC~\cite{Cepeda:2019klc} and at possible future colliders~\cite{Fujii:2017vwa,Roloff:2019crr,deBlas:2019rxi,Han:2020pif,Buttazzo:2020uzc,Bliewert:2024hed}. These improvements also benefit from the inclusion of the analysis of multi-scalar final states, see e.g. \cite{Stylianou:2023xit,Papaefstathiou:2023uum,Brigljevic:2024vuv} for analyses of triple-Higgs production.

Finally, another important property of the 125-GeV Higgs boson to verify is its decay width to invisible states --- specifically a pair of DM candidates $H$, when the decay channel $h\to HH$ is kinematically allowed. The current most stringent bound  $\text{BR}(h\to\text{invisible})\leq 10.7\,\%$ from ATLAS~\cite{ATLAS:2023tkt} implies that
\begin{equation}
\Gamma_{h\,\rightarrow\,H\,H}\,\leq\,\frac{0.107}{1-0.107}\,\Gamma_{h}^{\text{SM}}\,\approx\,0.120\,\Gamma_{h}^{\text{SM}}\,,
\end{equation}
where $\Gamma_h^\text{SM}$ denotes the 125-GeV Higgs boson width in the SM. In turn, this bound yields an upper limit on $\lambda_{345}$, which is related to the $hHH$ coupling. At the same time, for the mass range $M_H\leq M_h/2$, a lower value of $\lambda_{345}$ would lead to over-closing the Universe, because the cross-section of the DM annihilation process $HH\to h\to b\bar{b}$ would become too suppressed. One can therefore show~\cite{Ilnicka:2015jba,Belyaev:2016lok,Ilnicka:2018def,Dercks:2018wch,Kalinowski:2020rmb} that dark matter results together with the limits on $\text{BR}(h\to\text{invisible})$ produce a lower bound on the DM mass of $M_H\gtrsim 40\text{ GeV}$.

\section{Process generation and benchmarks}
\label{seq:procgen}

We concentrate on the VBF-type pair production of heavy scalars $A$ at a muon collider with a centre-of-mass energy of $10\text{ TeV}$. The specific target signal process that we consider is
\begin{equation}
    \label{eq:channel}
    \mu^+ \mu^- \rightarrow \nu_\mu \bar{\nu}_\mu A A \rightarrow \nu_\mu \bar{\nu}_\mu j j \ell^+ \ell^- H H,
\end{equation}
where the final state leptons $\ell$ can be either electrons or muons and the scalar $H$ is a stable dark-matter candidate contributing to missing energy. We consider the decay of the heavy scalar $A \rightarrow Z H$ with the two $Z$ bosons producing respectively a jet pair and a lepton pair. Background contamination arises from $\mu^+\mu^- \rightarrow j j \ell^+ \ell^- \nu \bar{\nu}$ with neutrinos of any flavour.  The presence of a heavy state contributing to the missing energy of the signal process results in distinctive kinematical characteristics that can be exploited to isolate signal events.

\subsection{IDM parameter scan}
By scanning over the parameter space of the IDM, we obtain points that fulfil the theoretical and experimental constraints described in \cref{sec:thexpconstraints}. In general a large mass splitting between the scalar $A$ and the dark matter candidate $H$ is kinematically favourable, leading to larger cross-sections for the target process. We note however that on the other hand mass splittings in the model are in general subject to several constraints such as perturbative unitary or bounds from electroweak precision observables. The mass splitting is correlated with the coupling $\bar{\lambda}_{345}$, defined in \cref{eq:l345b}, and the minimal values for the quantities entering this coupling are  $|\lambda_{345}| = 0$ and $M_H = 40$~GeV. The minimal scale for the dark matter candidate stems from a combination of signal strength and dark matter constraints, c.f.\ the discussion \cref{sec:thexpconstraints} and in Refs.~\cite{Ilnicka:2015jba,Kalinowski:2020rmb}.  

For this work, we combine several scan samples. As the most important parameters are given by the dark matter mass $M_H$, the upper scale of the additional new scalar masses, as well as the allowed values for $\lam_{345}$ after all constraints have been applied, we briefly list the ranges we considered for these quantities in \cref{tab:scans}.

\begin{table}[h]
    \begin{center}
    \begin{tabular}{l|c c c}
    Feature           & $M_H\,[\GeV]$ & $M_A,\,M_{H^\pm}$ $[\GeV]$ & $\lam_{345}$\\ \hline\hline
    Initial scan     & 55-100        & $\leq$ 500          & [-0.01;\;0.01] \\
    Large mass gap   & 110-200        & $\leq$ 640          & [-0.3;\;0.4]  \\
    Large mass range & $\leq$ 1000    & $\leq$ 1000         & [-0.02;\;0]
    \end{tabular}
    \caption{Different scan regions used within this work, characterised by the dark matter mass $M_H$, an upper mass scale (i.e. an upper limit on the considered ranges of $M_A$ and $M_{H^\pm}$), as well as allowed values of $\lam_{345}$ after all constraints are taken into account. \label{tab:scans}}
    \end{center}
\end{table}

For the first sample, we scan the parameter space for low-mass dark matter candidates, while we in general allow for masses up to 500 \GeV. In principle, such points generally lead to cross-section enhancements due to the available phase space. The second sample was designed to generate a large mass gap between the second inert neutral scalar $A$ and the dark matter candidate $H$, in order to enhance the respective coupling $\bar{\lam}_{345}$. In the third sample we allow in general for large masses of all inert scalars. For the selection of benchmark points, we make use of all three available samples.

\begin{figure}
    \centering
    \includegraphics[width=.7\textwidth]{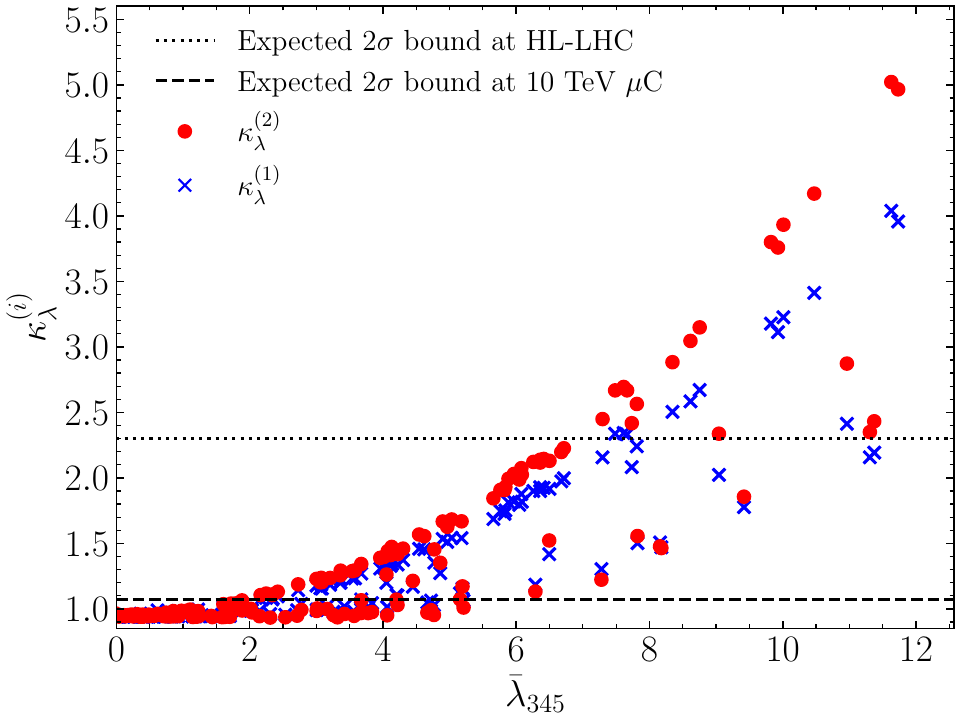}
    \caption{One- (blue crosses) and two-loop (red points) predictions for $\kappa_\lambda$ for our IDM parameter scan points. The vertical black lines indicate the expected experimental bounds on $\kappa_\lambda$, at the 95\% confidence level, from the HL-LHC (dotted) and from a 10-TeV muon collider (dashed). }
    \label{fig:kaplam_scatter}
\end{figure}

Before turning to the impact of $\bar{\lambda}_{345}$ on our target collider process, we note that this coupling can be constrained indirectly, via its correlation with corrections to Higgs properties like the trilinear Higgs coupling $\lambda_{hhh}$. At higher orders (i.e.\ from two loops and beyond), $\lambda_{hhh}$ depends on all the BSM parameters of the IDM, namely $M_H$, $M_A$, $M_{H^\pm}$, $\lambda_{345}$ and $\lambda_2$. However, the overall behaviour of $\lambda_{hhh}$ follows that of the dominant one-loop corrections involving the inert scalars. In turn, these are controlled by the couplings $\lambda_{345}$, $\bar{\lambda}_{345}$, and $\lambda_3$ --- which enter the $hHH$, $hAA$, $hH^+H^-$ interactions respectively. While for phenomenologically allowed points, dark matter data always constrain $\lambda_{345}$ to be small, large values of the coupling $\bar{\lambda}_{345}$ (or equivalently of the splitting between $M_H$ and $M_A$) are especially interesting for our target collider process and are usually correlated with a significant BSM deviation in $\lambda_{hhh}$ from radiative corrections involving $A$. This illustrated in \cref{fig:kaplam_scatter}, which displays one- (crosses) and two-loop (circles) predictions for the coupling modifier $\kappa_\lambda$ as a function of $\bar{\lambda}_{345}$ for the parameter scan points discussed above --- for the later, the leading two-loop corrections to $\lambda_{hhh}$ are computed using the results from Refs~\cite{Braathen:2019pxr,Braathen:2019zoh,Aiko:2023nqj}, while for both the one-loop corrections are obtained with the public code \texttt{anyH3}~\cite{Bahl:2023eau}. The results shown in this figure confirm that the general behaviour of $\kappa_\lambda$ with $\bar{\lambda}_{345}$ is relatively well-described by the one-loop corrections alone (i.e.\ the lowest order in perturbation theory at which contributions involving $\bar{\lambda}_{345}$ enter), while two-loop corrections can be numerically significant --- with relative magnitudes of up to $26\%$ of the one-loop contributions --- especially for large coupling values. It should be noted that the points with lower predictions for $\kappa_\lambda$ throughout the range of $\bar{\lambda}_{345}$ feature heavier masses of the dark matter candidate $H$ --- as can be seen from \cref{eq:IDM_0Lmasses}, this translates into higher values of the BSM mass scale $\mu_2$ (as $\lambda_{345}$ must remain small), and to heavier BSM scalars $A$ and $H^\pm$ for fixed $\bar{\lambda}_{345}$ and $\lambda_3$, which suppresses the corrections to $\lambda_{hhh}$. The dotted and dashed black vertical lines correspond to the $2\sigma$ expected limits on $\kappa_\lambda$ respectively from the HL-LHC~\cite{Cepeda:2019klc}, with $0.1<\kappa_\lambda<2.3$ at the 95\% CL using 3 ab$^{-1}$ of data, and from a high-energy muon collider, employing here the estimate of $\delta\lambda_{hhh}^\text{exp}=3.7\%$ at 68\% CL from Ref.~\cite{Accettura:2023ked}.
We can observe that the current constraints on $\kappa_\lambda$, discussed in \cref{sec:thexpconstraints}, do not reach the parameter space available from our scan. However, we can expect exclusion bounds on some of the model realisations for values of $\bar{\lam}_{345}$ down to about $\sim 7$ and $\sim 2$ for the HL-LHC and the muon collider, respectively. 

\subsection{Event generation and benchmark points}

Aiming to identify relevant observables for the target process in \cref{eq:channel} at a muon collider, we perform a numerical simulation generating events for both signal and background with \madgraph~\cite{Alwall:2014hca}. We use the {\textsc{UFO}}~\cite{Degrande:2011ua,Darme:2023jdn} model implementation of the IDM from Refs.~\cite{Goudelis:2013uca,Belanger:2015kga}
with the following input parameters, obtained from the Particle Data Group \cite{ParticleDataGroup:2022pth}
\begin{equation}
\begin{gathered}
\alpha_s=0.1179, \quad
(\alpha_\text{em})^{-1}=132.19,\quad
G_F=1.1663788\,\times\,10^{-5}\,\GeV^{-2}, \\ 
m_W=80.38\,\text{GeV},\quad
m_Z=91.188\,\text{GeV},\quad
m_h\,=\,125.3\,\GeV, \\ 
m_e=0.51099895\,\text{MeV},\quad
m_\mu=0.113428926\,\text{GeV},\quad
m_\tau=1.7769\,\text{GeV},\\
m_s=0.093\,\text{GeV},\quad
m_c=1.27\,\text{GeV},\quad
m_b=4.18\,\text{GeV},\quad
m_t=172.7\,\text{GeV}.  
\end{gathered}
\end{equation}
Additionally, we are using a non-diagonal CKM matrix in the Wolfenstein parametrisation, with the newest input parameters given by \cite{ParticleDataGroup:2022pth}\footnote{Note that the UFO model file available at \cite{repo} previously contained a faulty implementation of the CKM matrix, which however has recently been updated. We thank A.  Goudelis for useful discussions regarding this point.
}

\begin{equation}\label{eq:wstein}
    \lam\,=\,0.2250 (7),\;A=0.83 (2),\,\bar{\eta}\,=\,0.35 (1)\;\text{and}\;\bar{\rho}=0.16 (1)\;.
\end{equation}

For all cross-sections stated below that include the full final state, we are using the following pre-cuts at the generator level
\begin{equation}
\label{eq:cuts}
\begin{gathered}
p_T(j)\,\geq\,20\,\GeV,\quad
p_T(\ell)\,\geq\,10\,\GeV,\quad
\eta(j)\,\leq\,5.0,\\
\eta(\ell)\,\leq\,2.5,\quad
\Delta\,R_{i,j}\,\geq\,0.4,\,(i,j)\,\in\,\left[j,\ell\right].
\end{gathered}
\end{equation}
The background cross-section at $10$~TeV collisions for the electron (muon) final state with these pre-selection cuts is $3.47$ ($3.59$)~fb, while for $3$~TeV it reduces to $1.4$ ($2.1$)~fb.
 
Furthermore, we are generating the total widths for the new scalars for each parameter point explicitly via the three-body decays
\begin{equation}
    A\,\rightarrow\,H\,x\,y,\,\text{ and }\ H^\pm\,\rightarrow\,H\,x\,y,
\end{equation}
where $x,\,y$ 
denote stable SM final state particles. For the decay width of the 125-\GeV Higgs boson, we obtain the tree-level value
\begin{equation*}
\Gamma_h^\text{IDM,SM decs}\,=\,5.973 (1) \MeV
\end{equation*}
that includes both two- and three-body decays, as well as unstable electroweak gauge bosons in the final state. If non-vanishing, the partial decay width $\Gamma_{h\,\rightarrow\,H\,H}$ needs to be added to obtain a consistent total decay width.\footnote{Care must be taken when comparing this to the decay width that is given in the Yellow Report \cite{LHCHiggsCrossSectionWorkingGroup:2013rie, LHCHiggsCrossSectionWorkingGroup:2016ypw}. In the latter, the most up-to-date higher-order calculations are taken into account via the code \texttt{HDecay} \cite{Djouadi:1997yw,Djouadi:2018xqq}. In particular, the running of the masses is taken into account for quark decays, leading e.g.\ to the bottom mass $m_b\,\lb m_h\rb\,=\, 2.78859\,\GeV$ and therefore a modified leading order partial decay width of $\Gamma^\text{LO}_{b\,\bar{b}}\,=\,1.8894\,\MeV$. This discrepancy of roughly $2\,\MeV$ then is transferred also to the total width of the 125 \GeV particle, where \texttt{HDecay} gives 4.091 \MeV.} 

\begin{figure}[t!]
\begin{center}
    \includegraphics[width=0.7\textwidth]{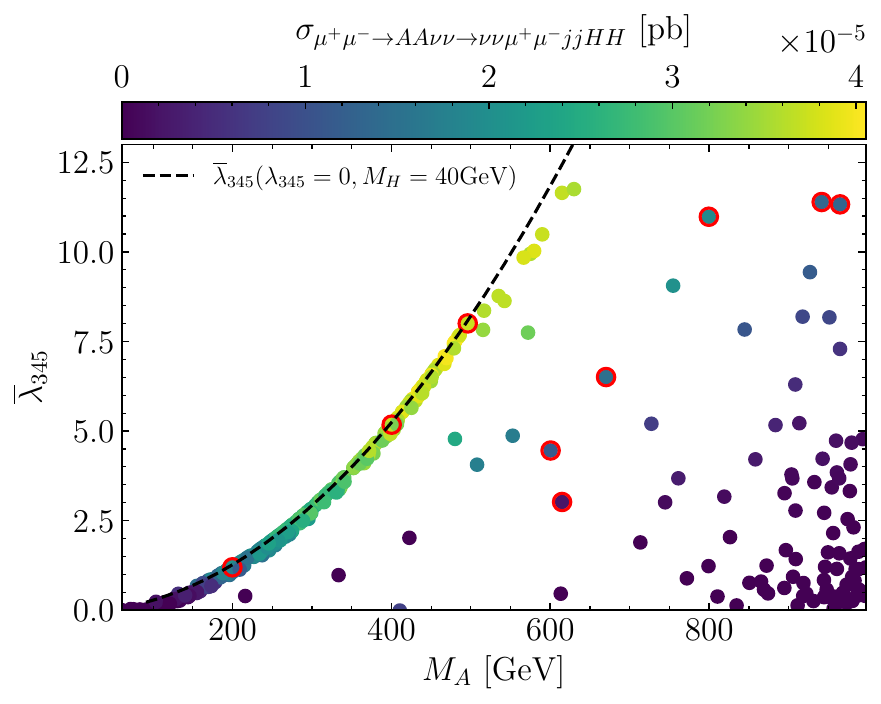}
    \caption{
    Total cross-section, with pre-cuts, for the channel \eqref{eq:channel} at $10$~TeV in the plane of $M_A$ and $\bar{\lambda}_{345}$. The dashed line corresponds to the maximal possible value of $\bar{\lambda}_{345}$ as a function of $M_A$, computed using \cref{eq:l345b} for $\lambda_{345}=0$ and $M_H=40\ \GeV$. Points lying on or near the dashed line come from the parameter scan with maximal $\bar{\lam}_{345}$ coupling. Red circled points mark the benchmark points defined in \cref{tab:bps}. 
    \label{fig:xsecs}}
    \end{center}
\end{figure}

Scalar masses in the interval $M_A\in \left[400, 600\right]$~GeV allow for a large mass splitting when $M_H$ is kept close to $40$~GeV and $\lambda_{345}$ is small, while also maintaining $\bar{\lambda}_{345} < 4 \pi $. For all scan points, all constraints described above are fulfilled.\footnote{During the completion of this work new results from the LUX-ZEPLIN collaboration for direct detection limits became available \cite{LZCollaboration:2024lux}. Two of the benchmark points we chose --- BP1 and BP4 --- are currently in tension with the corresponding bounds. In principle, these bounds can be avoided by setting $\lam_{345}$ to smaller values, which would only lead to sub-permille level changes in the corresponding production cross-sections, and thus not significantly modify our results for the sensitivities for these BPs at a 10 \TeV muon collider.} This parameter region yields the largest cross-sections for the channel under consideration. The cross-sections for the target process of the parameter scan points are displayed in \cref{fig:xsecs}.

Out of these parameter scan points, we select $11$ benchmark points (shown in \cref{tab:bps}) with various mass splittings to use for optimising our analysis and for comparisons. We additionally display the total decay widths of the scalar $A$ and the charged Higgs $H^\pm$, along with the cross-sections at $3$ and $10$~TeV collisions for these benchmark points in \cref{tab:xsecs}.

In \cref{tab:bps}, we also give the values for the NLO (i.e.\ two-loop) corrected di-photon decay rate, as well as the trilinear Higgs coupling modifier $\kappa_\lambda$ computed at two loops. From the bounds presented in  \cref{sec:thexpconstraints}, we observe that with the current central value for the di-photon rate, most points are in principle in disagreement with current measurements. However, we prefer to allow for a possible future shift of that rate to more SM-like values, which would then bring all points again into agreement assuming current uncertainties.

{\footnotesize
\begin{table}[h!]
\centering
\begin{tabular}{l c c c c c c c c}
\toprule
ID    & $M_H$  &    $M_A$  & $M_{H^\pm}$  &  $\lambda_2$  & $\lambda_{345}$ & $\bar{\lambda}_{345}$ &  $\Delta\kappa_{\gamma\gamma}^{(2)}$ [\%] & $\kappa_\lambda^{(2)}$ \\
\midrule
BP1   &  171.52  &  618.899  &  628.841     &  3.066190     &  0.14400 & 11.8307 & $-5.98$ & 5.22 \\
BP2    & 766.72  &  964.775  &  974.106     &  1.495400     &  $-0.00590$ & 11.3277 & $-2.56$ & 2.33\\ 
BP3  & 60.975  &  496.049  &  498.244     &  2.337340     &  $-0.00480$ & 8.00454 & $-5.93$ & 2.80\\
BP4  & 59.000  &  300.700  &  316.100     &  0.188496     &  $-0.00384$ & 2.86943 & $-5.39$ & 1.19\\
BP5  & 60.905  &  400.325  &  406.473     &  3.430620     &  0.00396 & 5.17782 & $-5.71$ & 1.73\\ 
BP6   & 62.400  &  199.800  &  230.000     &  0.138230     &  0.00486 & 1.19550 & $-5.16$ & 1.00\\ 
BP7   & 535.36  &  614.813  &  617.601     &  2.626370     &  $-0.00044$ & 3.01578 & $-1.42$ & 1.00\\ 
BP8   & 553.60  &  799.566  &  799.566     &  0.766550     &  $-0.01734$ & 10.9825 & $-3.36$ & 2.83 \\
BP9   & 474.88  &  600.384  &  618.238     &  3.593980     &  $-0.00328$ & 4.45670 & $-2.41$ & 1.21  \\
BP10  & 501.76  &  670.165  &  678.137     &  2.827430     &  $-0.01498$ & 6.50753 & $-2.74$ & 1.52 \\ 
BP11  & 736.00  &  941.656  &  947.464     &  0.942478     &  $-0.00926$ & 11.3933 & $-2.66$ & 2.41 \\
\bottomrule
\end{tabular}
    \caption{Definition of benchmark points used in this work. The last two rows show the two-loop predictions for $\Delta\kappa_{\gamma\gamma}^{(2)}$ and $\kappa_\lambda^{(2)}$ for these points. Dark matter variables are given in \cref{sec:dmprops}. 
    \label{tab:bps}}
\end{table}
}

\begin{table}[h!]
\centering
\begin{tabular}{llcccccc}
\toprule
ID & $\Gamma_A$ &       $\Gamma_{H^\pm}$ & $\sigma_e$ ($10$~TeV) &  $\sigma_e$ ($3$~TeV) &   $\sigma_\mu$ ($10$~TeV) &  $\sigma_\mu$ ($3$~TeV) \\
\midrule
BP1 & 56.83 &     63.41 & 3.80$\,\cdot\, 10^{-2}$ &     3.68$\,\cdot\, 10^{-3}$ & 3.80$\,\cdot\, 10^{-2}$ &     3.68$\,\cdot\, 10^{-3}$ \\
BP2 &    10.11 &     13.02 & 1.30$\,\cdot\, 10^{-2}$ &     1.51$\,\cdot\, 10^{-4}$ & 1.30$\,\cdot\, 10^{-2}$ &     1.52$\,\cdot\, 10^{-4}$ \\
BP3 &     34.43 &     36.79 & 3.71$\,\cdot\, 10^{-2}$ &     5.89$\,\cdot\, 10^{-3}$ & 3.71$\,\cdot\, 10^{-2}$ &     5.89$\,\cdot\, 10^{-3}$ \\
BP4 &      5.63 &      7.45 & 3.00$\,\cdot\, 10^{-2}$ &     8.88$\,\cdot\, 10^{-3}$ & 2.99$\,\cdot\, 10^{-2}$ &     8.88$\,\cdot\, 10^{-3}$ \\
BP5 &     16.36 &     18.29 & 3.55$\,\cdot\, 10^{-2}$ &     7.71$\,\cdot\, 10^{-3}$ & 3.55$\,\cdot\, 10^{-2}$ &     7.71$\,\cdot\, 10^{-3}$ \\
BP6 & 6.57$\,\cdot\, 10^{-1}$  &      1.89 & 2.14$\,\cdot\, 10^{-2}$ &     8.74$\,\cdot\, 10^{-3}$ & 2.13$\,\cdot\, 10^{-2}$ &     8.75$\,\cdot\, 10^{-3}$ \\
BP7 & 5.44$\,\cdot\, 10^{-3}$  & 2.65$\,\cdot\, 10^{-2}$  & 2.27$\,\cdot\, 10^{-3}$ &     1.97$\,\cdot\, 10^{-4}$ & 2.27$\,\cdot\, 10^{-3}$ &     1.97$\,\cdot\, 10^{-4}$ \\
BP8 &     18.58 &     19.92 & 1.93$\,\cdot\, 10^{-2}$ &     6.95$\,\cdot\, 10^{-4}$ & 1.93$\,\cdot\, 10^{-2}$ &     6.97$\,\cdot\, 10^{-4}$ \\
BP9 &      1.18 &      2.98 & 1.16$\,\cdot\, 10^{-2}$ &     1.03$\,\cdot\, 10^{-3}$ & 1.18$\,\cdot\, 10^{-2}$ &     1.03$\,\cdot\, 10^{-3}$ \\
BP10 &      4.86 &      6.61 & 1.48$\,\cdot\, 10^{-2}$ &     9.79$\,\cdot\, 10^{-4}$ & 1.48$\,\cdot\, 10^{-2}$ &     9.79$\,\cdot\, 10^{-4}$ \\
BP11 &     11.42 &     13.73 & 1.36$\,\cdot\, 10^{-2}$ &     1.89$\,\cdot\, 10^{-4}$ & 1.36$\,\cdot\, 10^{-2}$ &     1.89$\,\cdot\, 10^{-4}$ \\
\bottomrule
\end{tabular}
    \caption{Widths (in GeV) of the BSM particles and cross-sections (in fb) at $3$ and $10$~TeV for the benchmark points of \cref{tab:bps}. $\sigma_e$ ($\sigma_\mu$) denotes the cross-sections with electrons (muons) as the leptons in the final state. \label{tab:xsecs}}
\end{table}
    
\clearpage
\begin{figure}
    \centering
    \subcaptionbox{\label{fig:diaga}}{\includegraphics[width=0.3\textwidth]{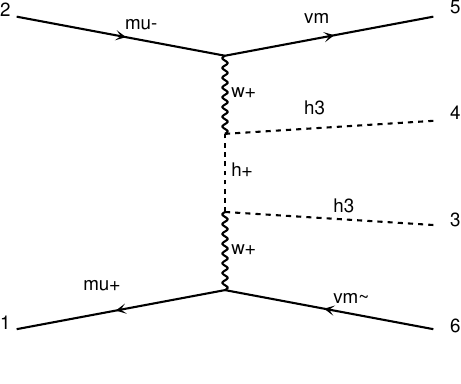}} \hspace{11mm}
    \subcaptionbox{\label{fig:diagb}}{\includegraphics[width=0.3\textwidth]{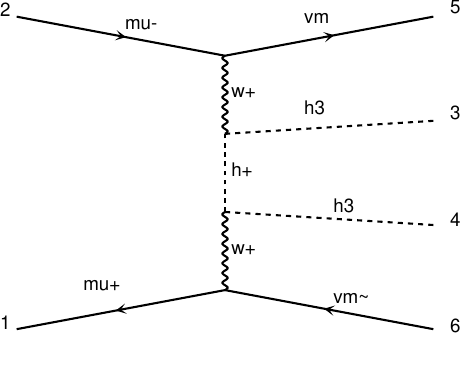}}\\
    \vspace{8mm}
    \subcaptionbox{\label{fig:diagc}}{\includegraphics[width=0.3\textwidth]{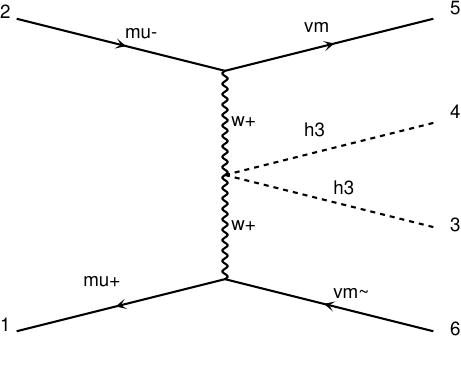}}\hspace{11mm}
    \subcaptionbox{\label{fig:diagd}}{\includegraphics[width=0.3\textwidth]{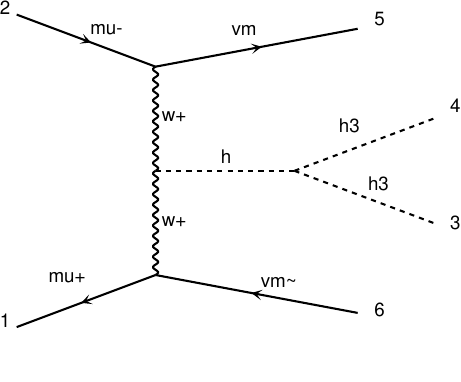}}\\
    \caption{Main topologies contributing to the target process \eqref{proc:main}. \textit{Top row and bottom left:} Processes governed by SM electroweak gauge boson couplings. These diagrams lead to large cancellations in the interference. \textit{Bottom right:} Additional diagram, $s$-channel contribution governed by $\bar{\lam}_{345}$. This is the dominant contribution after cancellation of the other diagrams. In these graphs, \texttt{h3} denotes $A$. This figure has been generated using \madgraph. }
    \label{fig:topos}
\end{figure}

\renewcommand{\arraystretch}{1.25}
\begin{table}[ht!]
        \centering
        \begin{tabular}{c|c c}
             \toprule
             &BP1&BP7\\ \hline
             $M_A \;[\GeV]$&618.899 &614.813 \\
             \rule{0cm}{.7cm}
             $\bar{\lam}_{345}$&11.8307&3.01578\\ \hline
             \rule{0cm}{.7cm}
             $\lvert (a) + (b) + (c) \rvert^2$ &0.00820(3)&0.00600(4)\\
             $\lvert (d) \rvert^2$ &2.19(2)&0.1468(7)\\
             $\lvert (a)  + (b)  +  (c) \rvert^2 + \lvert (d) \rvert^2$ &2.20(2)&0.1528(7)\\
             $\lvert (a) + (b) + (c) + (d) \rvert^2$ &2.37(2)&0.1641(7)\\ 
             All diagrams &2.35(1)&0.164(1) \\
             \bottomrule     
        \end{tabular}
        \caption{Important IDM parameters \textsl{(top rows)} and total integrated cross-sections, in \fb, for the process given by \cref{proc:main}, using subselections of diagrams from \cref{fig:topos}, as well as the total cross-sections without preselection of diagrams \textsl{(last row)}. We compare contributions for two benchmark points with similar masses for $A$ but different values for $\bar{\lam}_{345}$, namely BP1 and BP7. There are large cancellations between different diagrams governed by electroweak gauge couplings. The remaining cross-sections are proportional to $\bar{\lam}_{345}^2$. All cross-sections are shown for 10 \TeV. 
        }
        \label{tab:comp}
\end{table}
\renewcommand{\arraystretch}{1}

In order to understand the behaviour of the signal cross-section, it is instructive to consider different diagrammatic contributions to the target process, 
\begin{equation}\label{proc:main}
    \mu^+\,\mu^-\,\rightarrow\,A\,A\,\nu_\mu\,\bar{\nu}_\mu
\end{equation}
prior to decays. As an example, we compare BP1 and BP7, that feature similar masses for the scalar $A$ but largely different values for $\bar{\lam}_{345}$.

\begin{figure}[ht!]
\begin{center}
    \includegraphics[width=0.7\textwidth]{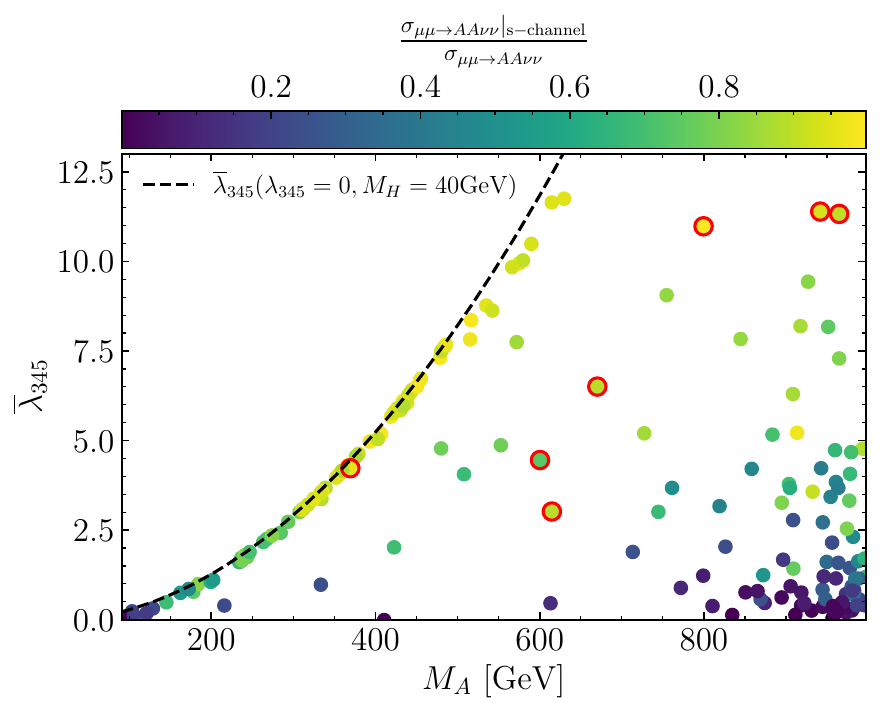}
    \caption{The ratio of the cross-section from $s$-channel diagrams in $\mu \mu \rightarrow A A \nu \nu$ divided by the total cross-section is shown. The contribution from the $s$-channel diagram (d) is dominating for larger values of $\bar{\lambda}_{345}$. The benchmark points contained in the considered random sub-sample are labelled with red circles. 
    \label{fig:scontri}}
    \end{center}
\end{figure}

In \cref{fig:topos}, we display dominant contributions from VBF-type topologies, which we denote (a), (b), (c), and (d). It is known that there are large cancellations for different contributions that are related via unitarity considerations. In our scenario, we therefore group processes that stem from VBF-type topologies and are governed by SM electroweak gauge couplings. For the process at hand, these are VBF processes that have a charged scalar in the $t$-channel --- c.f.\ (a) and (b) in \cref{fig:topos} --- as well as the VBF topology with the quartic $A\,A\,W^{+}\,W^{-}$ coupling --- diagram (c) in \cref{fig:topos}. In these processes, the couplings are determined by SM parameters, and the only new physics parameter that enters is the mass $M_A$ (at leading order). \cref{tab:comp} shows the different contributions from these channels for the two benchmark points under consideration. In addition, we list the process where an $s$-channel SM-like Higgs boson mediates the process --- i.e.\ diagram (d) in \cref{fig:topos} --- as well as the sum with and without interference. We see that there is an additional interference effect of around $7\%$ for both considered scenarios. We equally in the last row display the total cross-section using all diagrams that are generated by \madgraph~ for this process. In addition, for diagram (d) the ansatz 
\begin{equation*}
    \sigma_{(d)}\,\sim\,\bar{\lam}_{345}^2
\end{equation*}
is found to be satisfied at the percent level.

One can furthermore investigate which contribution of the total cross-section stems from the diagram with an $s$-channel Higgs-like particle, given by diagram (d) in \cref{fig:topos}. We show the percentage in colour coding of this contribution with respect to the total cross-section in \cref{fig:scontri} for a smaller subset of randomly chosen scan points. The contributions from $s$-channel diagrams are dominating for larger values of $\bar{\lambda}_{345}$.

As we later plan to compare with results achievable at a centre-of-mass energy of 3 \TeV, we already comment on this setup here. Indeed, a similar investigation for a centre-of-mass energy of 3 \TeV gives results similar to those shown in \cref{tab:comp}; we find the above diagrams to be dominant, with large cancellations between contributions governed by electroweak gauge couplings, as well as a proportionality of diagram (d) contributions to $\bar{\lam}_{345}$. However, the total cross-sections are reduced by about one order of magnitude.

We can also check the impact of different masses with a similar value for $\bar{\lam}_{345}$. BP1 and BP11 have a similarly large value for this coupling, but different masses $M_A$. A calculation of the $s$-channel contribution alone at 10 \TeV reveals that an increase from $M_A\,\sim\,620\,\GeV$ to $\sim\,940\,\GeV$ leads to a decrease of the $\mu^+\mu^-\to\nu_\mu\bar{\nu}_\mu AA$ cross-section (before cuts) by roughly a factor 4.4. 

Additionally, we investigate the contribution of different channels for the benchmark point with the smallest value of $\bar{\lam}_{345}$, BP6. For this point we find that the contributions from the subprocesses mediated by electroweak gauge bosons amount to roughly $9\%$ of the total cross-section. The $s$-channel mediated diagram still contributes about $56\%$, followed by $34\%$ from the interference terms. We note that for all discussions above, we have considered contributions in the unitary gauge.

\begin{figure}[t!]
    \centering
    \includegraphics[width=0.49\textwidth]{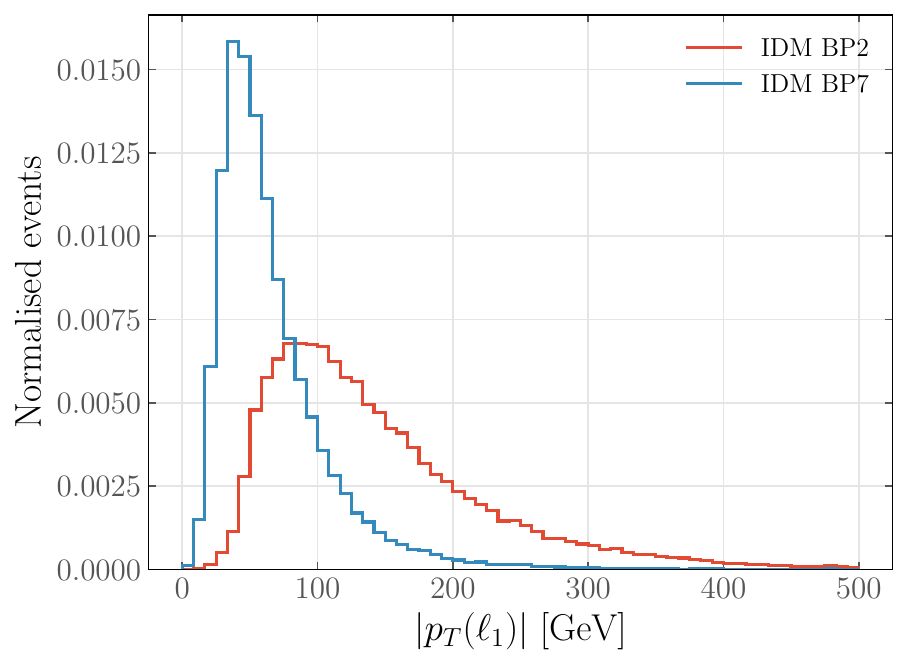}
    \includegraphics[width=0.49\textwidth]{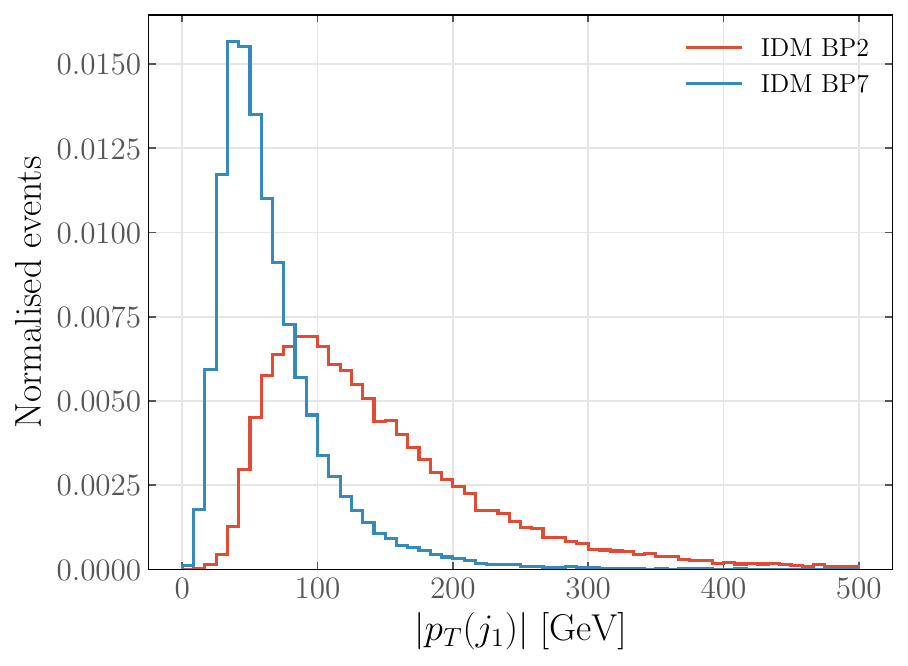}
    \caption{Transverse momentum distributions for the leading lepton and jet for BP2 (red) and BP7 (blue), normalised to unity.}
    \label{fig:pt}
\end{figure}

Finally, the pre-cuts listed in \cref{eq:cuts} also have a different impact on the different benchmark points. As an example, we discuss BP7 and BP2. These two parameter points differ both in the absolute scale of the inert scalar masses, given basically by $M_H$, as well as the mass difference $M_A-M_H$. In particular, for BP7 this difference is around 80 \GeV, resulting in softer decay products. For BP7 and BP2, the pre-cuts lead to a reduction to $38\%$ and $73\%$, respectively, of the total cross-section without cuts. This is mainly due to the different kinematics of the BPs. As an example, we therefore display the $p_T$ distributions of the leading $p_T$ lepton and jet for these two processes in \cref{fig:pt}, which we normalise to unity. In particular the cut on $p_T(j)\geq 20\ \GeV$ has a different impact on the two benchmark points, leading to a reduction of the cross-sections to $60\%$ and $89\%$, respectively. Transverse momentum cuts reduce the cross-sections to $52\%$ and $86\%$ of the original rate for BP7 and BP2, respectively.

\subsection{Signal contributions from additional processes}
Additional channels could lead to the same signature as our target process. In principle, this signature can be reached via production of two scalars $A$ decaying as in the channel~\eqref{eq:channel} but with or without additional neutrinos or invisible scalars $H$. Furthermore the IDM charged scalars $H^\pm$ decaying to $W^\pm H$ could also provide further contributions to the same final state. We have therefore also investigated the following processes:
\begin{itemize}
    \item production of two scalars $A$ through $s$-channel diagrams (without additional neutrinos) ,
    \begin{equation*}
    \mu^+ \mu^- \rightarrow (A \rightarrow \ell^+ \ell^- H) (A \rightarrow j j H)\;,
    \end{equation*}
    \item production of two BSM scalars $H$, which contribute to the missing energy, and two scalars $A$,
    \begin{equation*}
    \mu^+ \mu^- \rightarrow H H (A \rightarrow \ell^+ \ell^- H) (A \rightarrow j j H)\;,
    \end{equation*}
    \item production of two scalars $A$ and two non-muon neutrinos,
    \begin{equation*}
        \mu^+ \mu^- \rightarrow \nu_{e,\tau} \bar{\nu}_{e, \tau} (A \rightarrow \ell^+ \ell^- H) (A \rightarrow j j H)\;,
    \end{equation*}
    \item and production of two charged scalars $H^\pm$,
    \begin{equation*}
        \mu^+ \mu^- \rightarrow (H^\pm \rightarrow \ell^\pm \pbar{\nu} H) (H^\mp \rightarrow j j \ell^\mp \pbar{\nu} H)\;.
    \end{equation*}
\end{itemize}

We calculate the cross-sections for these additional channels for BP1, BP2 and BP3,
using the same generation-level cuts as the main channel for $10$~TeV collisions. However, our findings indicate that all channels are negligible, since the contributions are smaller than $1\%$ of the cross-sections in \cref{tab:xsecs}. The largest contribution would be from the production of $H^\pm H^\mp$ (i.e.\ the last process in the list above) which is still subdominant. Thus, we do not include them for our analysis in \cref{sec:analysis}. It should nevertheless be noted that these additional processes would only enhance the rates of signal events leading to higher statistical significances. 
\section{Analysis}
\label{sec:analysis}

Evaluating the potential of a muon collider to uncover effects arising from IDM states requires well-optimised signal-background discrimination to separate the BSM contributions from the SM. On the one hand, cut-and-count analyses defining the signal region with cuts on appropriate observables are simple and rather intuitive. On the other hand, ML techniques can greatly enhance the significance of a particular channel by setting highly sophisticated non-rectangular cuts that define the optimal signal region. This is, however, done at the cost of interpretability, as it is particularly hard to identify  which kinematical quantities allow obtaining a higher sensitivity. 

The challenge in discriminating the signal from the background arises from the presence of multiple particles in the final states and the fact that the missing energy is due to both neutrinos and scalars $H$.  We design two different analyses, one with traditional cuts and one with Boosted Decision Trees (BDTs), aiming to identify the relevant observables for the process \eqref{eq:channel}. For the latter case we investigate which observables are crucial in order to separate the background from the signal using Shapley values~\cite{shapley}, see \cref{sec:shap}. 

For both approaches we consider muon collisions at $10$~TeV with an integrated luminosity of 10 ab$^{-1}$, following the relation~\cite{Accettura:2023ked}\footnote{Luminosity targets were updated in \cite{MuCoL:2024oxj}. For 3 \TeV, there is a $10\%$ discrepancy between the cited luminosities. We do not expect this to have a significant impact on the results of our study.}
\begin{equation}
    \label{eq:lumi}
   {\mathcal{L}}_\text{int} \,=\,10\, \text{ab}^{-1} \left(\frac{E_\text{CM}}{10\text{ TeV}}\right)^2 \,.
\end{equation}

\subsection{Cut-and-count analysis}

\begin{figure}
    \centering
    \includegraphics[width=.49\textwidth]{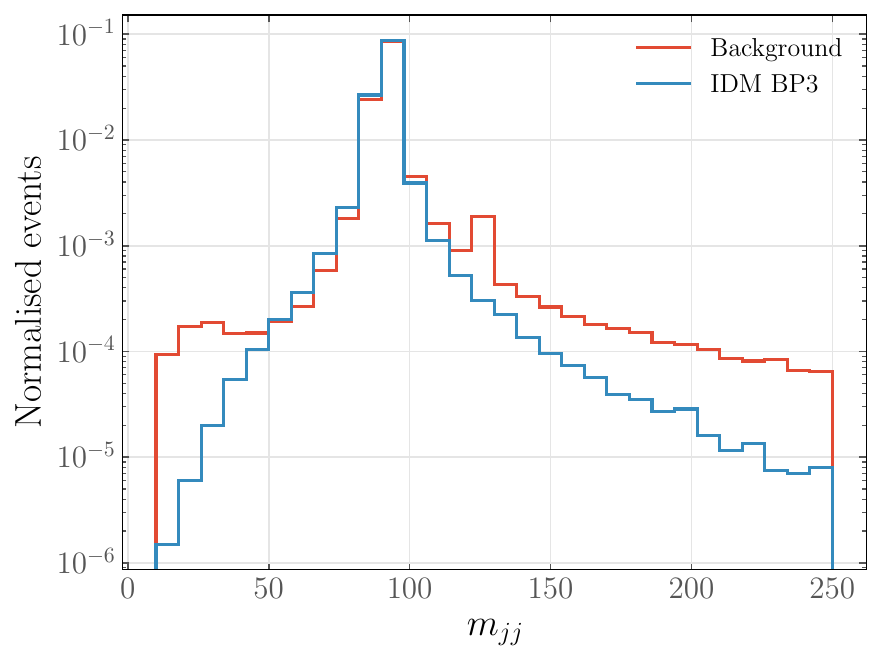}
    \includegraphics[width=.49\textwidth]{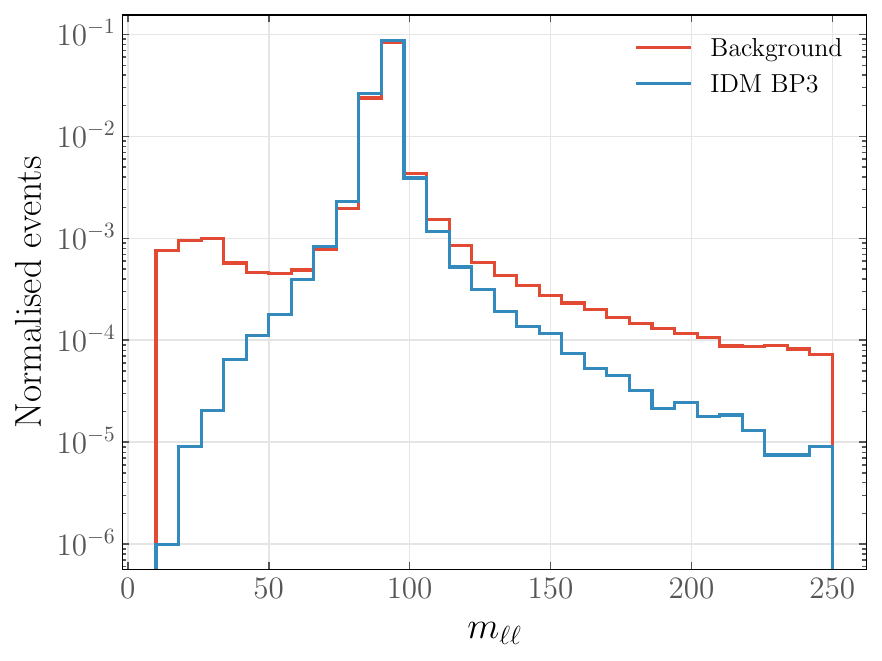}\\
    \includegraphics[width=.49\textwidth]{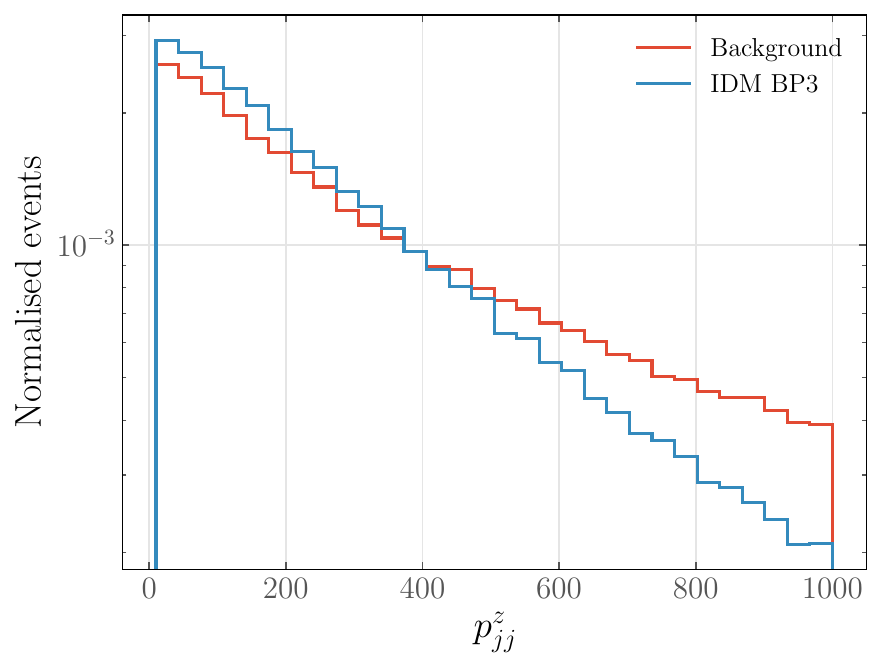}
    \includegraphics[width=.49\textwidth]{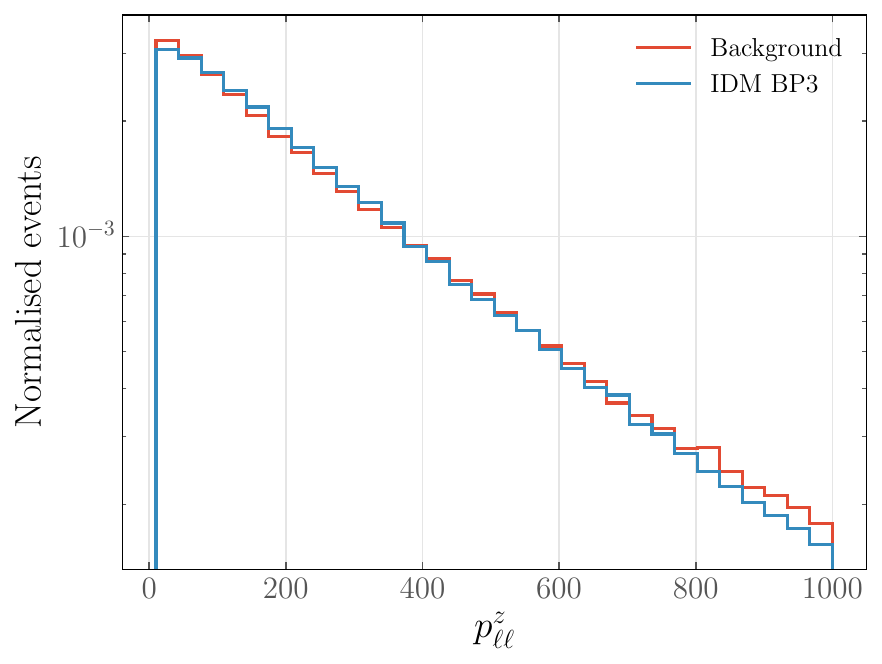}\\
    \includegraphics[width=.49\textwidth]{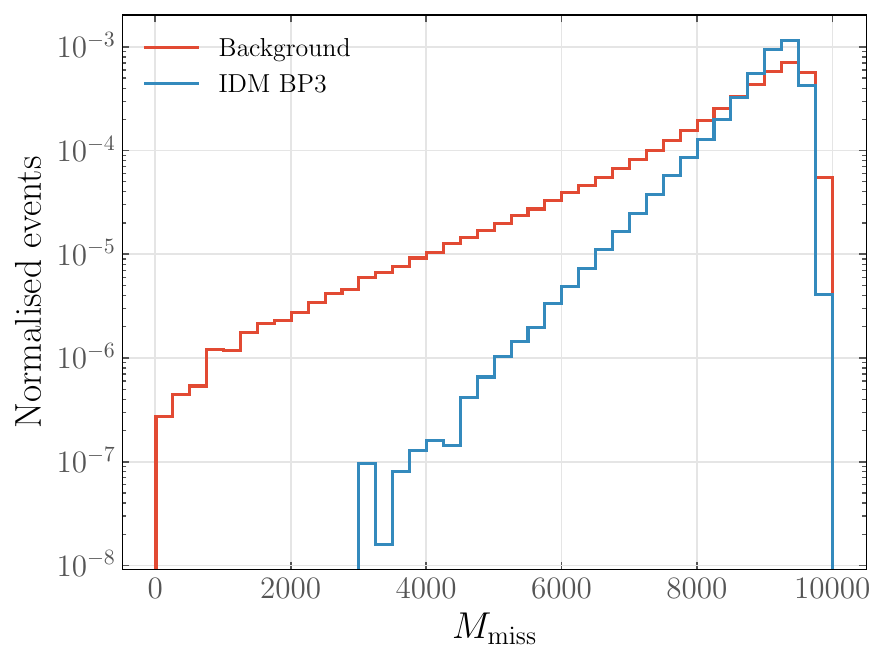}
    \includegraphics[width=.49\textwidth]{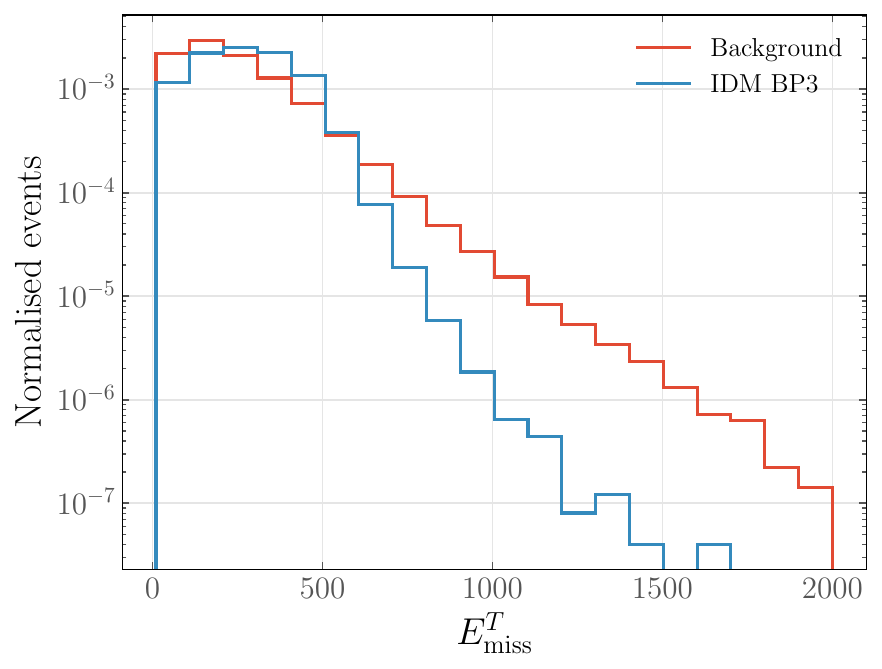}
    \caption{Distributions of the most important observables utilised in both the cut-and-count and ML analyses for the background (red) and for BP3 (blue).} 
    \label{fig:histos}
\end{figure}

The cut-and-count analysis serves the purpose of a baseline indicating the potential prospects of the muon collider and also allows quantifying the gain from using ML techniques. We optimise our cuts for BP3, which has a relatively large cross-section at $10$~TeV collisions, and subsequently use the same cuts for all parameter points. For this benchmark point, we present histograms for the observables with discriminating power in \cref{fig:histos}.

\edit{We note that while for some of the displayed variables --- as e.g.\ the invariant mass distributions --- signal and background show similar shapes in particular in the dominant regions, variables that include the missing momenta, as e.g.\ $M_{\text{miss}}$ or $E^T_{\text{miss}}$ exhibit significant differences in particular in the region of low missing mass and high missing transverse energy. We therefore take these as variables that can be used to discriminate signal from background in this particular benchmark scenario.} 

Since the signal process proceeds via decays of $A \rightarrow Z H$, a large contribution to the signal comes from the case when the $Z$ boson is on-shell and it is thus beneficial to remove contributions from the background where the visible final states do not arise from a $Z$ resonance. We impose a cut on the invariant mass of the di-jet $jj$ and di-lepton $\ell\ell$ states, $40 < m_{jj/\ell\ell} < 130\text{ GeV}$ to enforce this. The region of interest can be further restricted with $p^z_{jj/\ell\ell} < 900\text{ GeV}$, where $p^z_{jj/\ell\ell}$ is the $z$-component of the pair's four-momentum.  While both signal and background have invisible particles leading to large missing momentum, the presence of heavy particles among the missing energy final states of the signal leads to differences in the distributions of related observables. In particular, we define the missing four-momentum as  ${p}_\text{miss} = \left(\sqrt{s}, \vec{0}\right) - \sum p_\text{vis.}$ where $p_\text{vis.}$ are the four-momenta of the jets and leptons. The missing invariant mass $M_\text{miss} = \sqrt{{p}_\text{miss} \cdot p_\text{miss}}$ is then an important observable for discriminating purposes as the majority of signal events attain larger values of $M_\text{miss}$ than the background. We therefore set a cut of $M_\text{miss} > 8250$~GeV and also impose a cut on the missing transverse energy $E^T_\text{miss} = \sqrt{{p}_{\text{miss}, x}^2 + {p}_{\text{miss}, y}^2} < 2000\text{ GeV}$. These cuts have been identified by scanning over different possible values and identifying the thresholds that maximise the significance for our channel. The cutflow of the aforementioned cuts is shown in \cref{tab:cutflow}. The majority of the signal events is retained; however given the relatively large acceptance of background events after the cuts it is also apparent that the simple rectangular cuts do not drastically reduce the background contribution.\footnote{We also have considered other observables such as the stransverse-mass $m_{T2}$~\cite{Lester:1999tx} and angular quantities (e.g. differences of pseudorapidities and azimuthal angles of the lepton and jet pairs). However, they are not particularly useful for reducing the background in this channel.}
    
We sum the number of signal and background events that we observe in our signal region for the muon and electron final states and subsequently evaluate the significance for each parameter point with the formula \cite{Cowan:2010js} 

\begin{equation}
    \label{eq:significance}
    Z = \sqrt{2 \left( \lb S + B\rb \log\left[ 1 + \frac{S}{B}\right] - S \right) } \;,
\end{equation}
where $S$, $B$ denote the signal and background, respectively. \edit{
As customary, we here define the expected discovery significance as the expected $Z$ under the assumption of some nominal signal model. 
}

\begin{table}[h!]
\begin{center}\begin{tabular}{lcccc}
\toprule
{} & BP3 ($e$) &  Background ($e$) & BP3 ($\mu$) &  Background ($\mu$) \\
\midrule
$40 < m_{jj} < 130$ \GeV      &    0.993 &    0.975 &    0.993 &    0.976 \\
$p^z_{jj} < 900$ \GeV         &    0.943 &    0.824 &    0.943 &    0.827 \\
$40 < m_{\ell\ell} < 130$ \GeV &    0.936 &    0.768 &    0.936 &    0.786 \\
$p^z_{\ell\ell} < 900$ \GeV    &    0.911 &    0.730 &    0.911 &    0.748 \\
$M_\text{miss} > 8250$ \GeV        &    0.862 &    0.608 &    0.861 &    0.628 \\
$E^{T}_\text{miss} < 2000$ \GeV  &    0.862 &    0.608 &    0.861 &    0.628 \\
\bottomrule
\end{tabular}
\end{center}
\caption{Reduction of signal and background acceptances from the cut-and-count analysis cutflow for the electron and muon cases. \label{tab:cutflow}}
\end{table}

The resulting significances for the benchmark points under consideration are shown in the left plot of \cref{fig:bp_signif}. The largest values of $\sim 3$ are obtained for BP1, BP3 and BP5 and are characterised by a substantial mass splitting between $M_H$ and $M_A$. Smaller mass differences are associated with a reduced sensitivity and for higher values of $M_A$ the significance starts again to drop. While the cut-and-count analysis does show promise of finding evidence of phenomena arising from the IDM, a more optimal selection of the signal region with modern ML techniques closer to what experiments are currently utilising would yield better results. We therefore investigate this using BDTs in the next section.

\subsection{ML analysis}
While it is impossible to anticipate how advanced the performance of ML techniques will be at the time scale of a muon collider, it would presumably surpass current techniques. BDTs are amongst the most popular ML techniques utilised in High-Energy Physics, able to yield improved results for various tasks ranging from triggers at experiments to event-level classification of different contributions (for an exhaustive review see Ref.~\cite{Feickert:2021ajf}). We therefore choose to utilise gradient BDTs to showcase what significance a muon collider would likely reach. We use the \xgboost~\cite{xgboost} library, interfaced through \sklearn~\cite{scikit-learn}, in order separate the signal and background IDM contributions at the $10$~TeV muon collider. For each benchmark point under consideration, we train a separate \xgboost~tree which is used to identify the signal region and train different BDTs for the electron and the muon final states.  

The loss function used during training is the log-likelihood of the Bernoulli distribution and the step size shrinkage~\cite{xgboost} (or learning rate) is set to $0.034$. The maximum number of trees in the ensemble is $267$ and the depth of each tree can not exceed $11$. We fix the minimum sum of the weight instances to $7$. For each event, the following observables are included:
\begin{itemize}
    \item the pseudorapidities $\eta$ and transverse momenta $p_T$ of the jets and leptons,
    \item the azimuthal angle difference between the jets (leptons) $\Delta\phi_{jj}$ ($\Delta\phi_{\ell\ell}$) and the pseudorapidity difference $\Delta\eta_{jj}$ ($\Delta\eta_{\ell\ell}$),
    \item the invariant dilepton and dijet masses $m_{\ell\ell}$ and $m_{jj}$, respectively, as well as their momenta along the $z$-direction $p^z_{\ell\ell}$, $p^z_{jj}$,
    \item the missing invariant mass $\mmiss$ and the missing transverse energy $\etmiss$.
\end{itemize}
A total of 200,000 signal and 200,000 background events are considered for each benchmark point and each final state. We retain $10\%$ of the events to evaluate the performance of the algorithm and train on the rest.

\begin{figure}[t!]
    \centering
    \includegraphics[width=.49\textwidth]{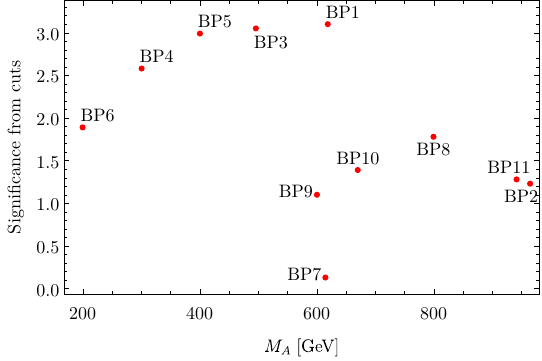}
    \includegraphics[width=.49\textwidth]{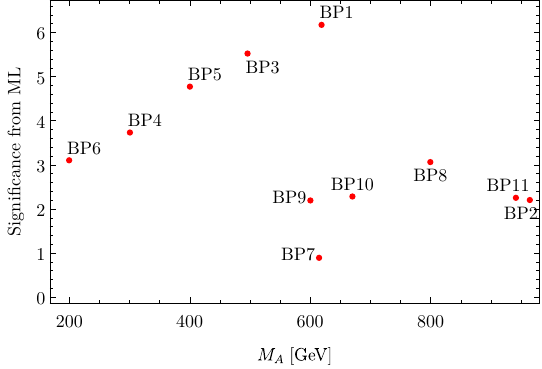}
    \caption{Significance using cuts (left) and ML (right) for the benchmark points of \cref{tab:bps}. }
    \label{fig:bp_signif}
\end{figure}

Signal and background efficiencies are calculated for different thresholds of the {\xgboost} output score for each trained tree. By identifying the working point yielding the highest significance, the signal region is algorithmically defined and we extract the number of signal and number of background events for the electron and muon final state channels. Subsequently, we add the events from the two processes and evaluate the final significance for each benchmark point with the expression in \cref{eq:significance}. We show the significances for our identified benchmark points in the right plot of \cref{fig:bp_signif}. In comparison with the cut-and-count analysis, the ML approach expectedly achieves higher sensitivity, reaching significances of $\sim 6$ for BP1. The overall pattern remains similar, with larger mass differences between the dark-matter candidate $H$ and the scalar $A$ resulting in larger sensitivity. The fact that large mass differences lead to large sensitivity is related to the increased cross-section for these points, exactly like the cut-and-count analysis. However, the ML method is insensitive to the cross-section itself, it only defines the appropriate signal region (in other words we do not use the cross-section as an input to the BDT). Therefore, the increase of the significance compared to the cut-and-count analysis mainly stems from kinematics.

\begin{figure}[h!]
    \centering
    \includegraphics[width=.49\textwidth]{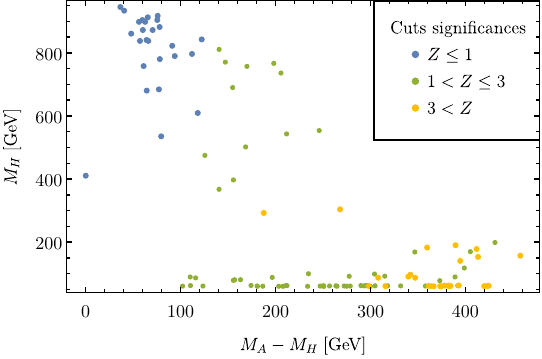}
    \includegraphics[width=.49\textwidth]{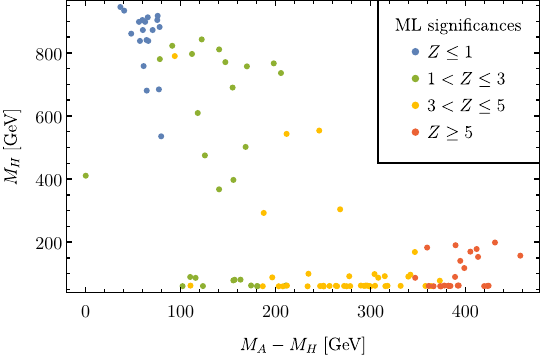}
    \caption{Subset of parameter points that were used for the cut-and-count analysis (left) and the ML analysis (right) shown on the $M_A - M_H$ vs.~$M_H$ plane. }
    \label{fig:full_scan_signif}
\end{figure}

\begin{figure}[h!]
    \centering
    \includegraphics[width=.6\textwidth]{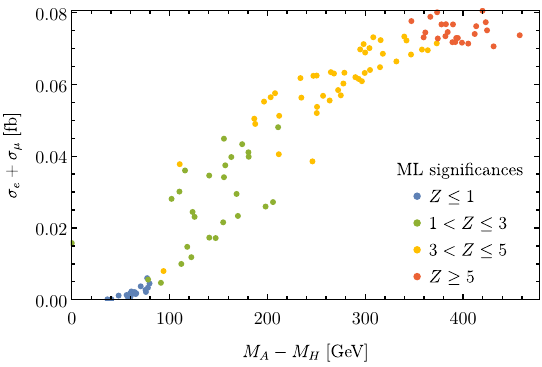}
    \caption{Cross-sections after pre-cuts defined in \cref{eq:cuts} for the same subset of parameter scan points as in \cref{fig:full_scan_signif} as a function of $M_A - M_H$, and along with the ML significance. Large mass differences are associated with enhanced cross-sections and yield a higher significance. 
    \label{fig:full_scan_signif_xsec}
    }
\end{figure}

Finally, we apply our cut-and-count and ML analyses to a larger set of parameter points allowed by theoretical and experimental constraints. The results are shown in \cref{fig:full_scan_signif} and indicate that a large statistical significance is expected for points characterised by a small $M_H$ value but also a large mass separation $M_A - M_H$. As expected by the previous discussion, the ML approach outperforms the cut-and-count analysis yielding significances larger than $Z = 5$ (red points). \cref{fig:full_scan_signif_xsec} showcases the correlation between large mass differences and larger cross-sections that leads to an enhanced significance. One can roughly state that production cross-sections $\geq\, 0.07\, (0.05)\, \fb$ lead to significances $\geq\, 5\, (3)$ using our methodology.

Aiming to interpret the behaviour of our ML approach and enhance our understanding of the results, we have utilised Shapley values~\cite{shapley} in~\cref{sec:shap}.

\begin{figure}
    \centering
    \includegraphics[width=0.49\textwidth]{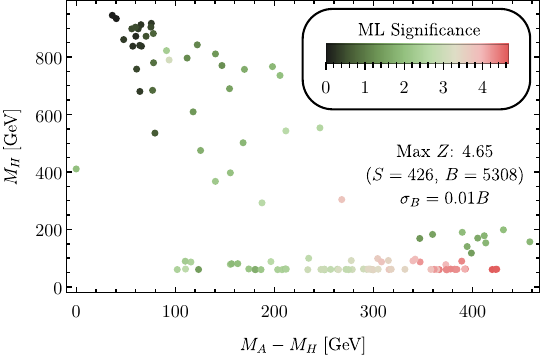}
    \includegraphics[width=0.49\textwidth]{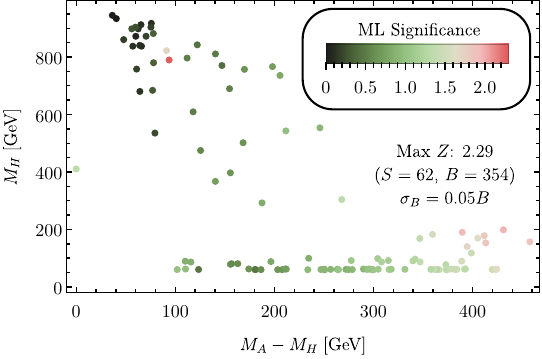}
    \caption{
    \edit{
    The significance for each parameter point with an assumed $\sigma_B = 1\% B$ and $\sigma_B = 5\% B$ systematic uncertainty is shown on the left and right plot, respectively. We additionally indicate the maximum significance for each case along with the corresponding number of signal and background events for the particular parameter point. 
    }
    \label{fig:significances_with_syst}
    }
\end{figure}

\edit{The technical development of a muon collider with a 10 \TeV center of mass energy is currently ongoing. Therefore, clear estimates of systematic uncertainties are not yet available. Assuming these to be small due to the relatively clean environment of a lepton collider, we briefly investigate how the significance would change if such uncertainties were taken into account. We use the expression \cite{Cowan:2010js,slac12,medsig}}
\begin{equation}
\label{eq:syst}
Z\lb S,B\rb\,=\sqrt{2\Biggl(\Bigl[S+B\Bigl] \ln\Biggl[\frac{(S+B)(B + \sigma^2_B)}{B^2+ (S + B)\sigma^2_B}\Biggl]-\frac{B^2}{\sigma^2_B}\ln\Biggl[1 + \frac{\sigma^2_B S}{B(B +\sigma^2_B)}\Biggl] \Biggl)}\,,  
\end{equation}
\edit{where $\sigma_B$ is an estimate of the background systematic uncertainty. 
The impact for systematic uncertainties equal to $1\%$ and $5\%$ background events is shown in \cref{fig:significances_with_syst} along with the highest significance in each case and the corresponding signal and background events. With $1\%$ systematic uncertainties the maximum significance is slightly below the discovery limit, while for $5\%$ the significance of all parameter points is substantially reduced. This is due to the large number of background events compared to signal events for all the parameter points. Signal events after the ML selection are in the range $S \in [\![1,426]\!]$, while background events are considerably higher $B \in [\![354,10521]\!]$. The precise determination of the background is therefore crucial for the scenario under investigation.
}

\subsection{Comparison with $3$~TeV collisions}
\begin{figure}[t!]
    \centering
    \includegraphics[]{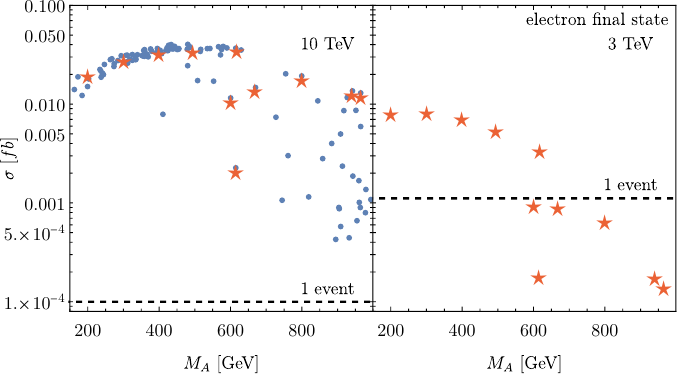}
    \caption{Cross-sections for the production cross-section of electron final states for all parameter points used in our study are shown on the left for a 10~TeV muon collider. The starred points correspond to the benchmark points of \cref{tab:bps} which are also shown for 3~TeV collisions on the right. 
    \label{fig:xsec_10vs3}}
\end{figure}

In order to understand whether there is sufficient motivation to increase the energy to $10$ \TeV, we revisit the case of $3$ \TeV muon collisions with a luminosity of $\sim 0.9\text{ ab}^{-1}$, obtained from \cref{eq:lumi} (we note that if we used instead as target luminosity at 3 TeV the value of 1 ab$^{-1}$ from e.g.\ Ref.~\cite{Accettura:2024qcg}, our results would not be drastically modified). The cross-sections for the benchmark points are shown in~\cref{fig:xsec_10vs3}, where we include additional parameter points from the scans  for the $10$ \TeV case. We display case of electrons in the final state as an example, with similar values attainable for the case of muons.

A few comments are in order regarding the cross-sections in this figure. First, as already discussed above, the total cross-sections depend both on the scale of the process, set by $M_A$, as well as on the value of $\bar{\lam}_{345}$. In addition, kinematics differ between the benchmark points, leading to different cut efficiencies for the precuts. Independently of this, we found that \cref{fig:xsec_10vs3} displays well the overall behaviour of the total cross-sections. In particular, we see that the energy that is available to the initial electroweak gauge bosons that enter into the VBF-type topology and constitute the dominant contributions for our target process largely differs for the different centre-of-mass energies.

We can investigate this behaviour for BP1, which features the largest difference in production cross-sections prior to decays between the two different centre-of-mass energies (the comparison is done normalising the cross-sections to that of BP6 at 10 \TeV). In order to simulate the events on a similar level, we decide to include an invariant mass cut of $m_{AA}\leq3\,\TeV$ in both cases for this comparison. The total production cross-sections for these two centre-of-mass energies before decays are $\sim\,2.2\,\fb$ and $\sim\,0.18\,\fb$ for 10 \TeV and 3 \TeV, respectively. As the dependence of matrix element for $W^+\,W^-\rightarrow A\,A$ on the actual parton-level centre-of-mass energy for this subprocess is unaffected by the total centre-of-mass energy of the collider, this change of cross-section for a given benchmark point by an order of magnitude can only be attributed to the available energy of the $W$ bosons, that are governed by the so-called ``$W$ content of the muon'' (see e.g. \cite{Han:2020uid,Garosi:2023bvq,Ciafaloni:2024alq} for details). In addition to this, the estimated total integrated luminosity for a 3 \TeV machine is lower by about one order of magnitude.

As the background rate is also reduced, one may wonder whether it might be possible to have enough sensitivity already at lower-energy collisions. Therefore, we follow the same procedure as for the $10$ \TeV ML analysis, generating signal and background samples of the same size. We train new BDTs using the same hyperparameters and identify thresholds on the ML output score that maximise the significance as before. 

Overall, our findings after combining the two final states as before indicate that muon collisions at $3$~TeV will be unable to uncover signs of the IDM in this channel. The highest significance of $0.53$ is achieved for BP6, which highlights the limited sensitivity. While the cross-section rates are not reduced that drastically, the analysis suffers from the notably reduced integrated luminosity. With the current target luminosity, see \cref{eq:lumi}, we conclude that a 10 \TeV muon collider is needed to explore the parameter space of the IDM considered here.



\section{Conclusion}
\label{sec:conclusions}

In this work, we discussed the discovery potential of a 10 \TeV muon collider for novel scalar production in the Inert Doublet Model using VBF-type production modes. We briefly presented the current state-of-the-art constraints on the parameter space of the model, including higher-order corrections to the trilinear Higgs coupling and Higgs to di-photon decay width. 
We investigated in detail the mechanism for $A\,A$ production including its dependence on the BSM input parameters, and performed a comparison of the 10 \TeV collider case to one with a centre-of-mass energy of 3 \TeV.  After including relevant background processes, we found that relatively high significances can be achieved using both cut-based as well as ML-improved methods, where the latter surpasses the former in terms of the discovery potential. We can set approximate lower limits on production sections of about $0.07\,(0.05)\,\fb$ needed for discovery or exclusion, respectively, applying our ML methodology. Furthermore, we found the highest significances for low DM masses $M_H\, \lesssim\, 200\,\GeV$ in combination with mass gaps $M_A-M_H$ around 400 \GeV. While for a number of the benchmark points we considered, measurements of properties of the 125-GeV Higgs boson could in principle exhibit deviations from SM predictions by the time a muon collider would be built --- thereby providing indirect evidence for BSM Physics --- the direct searches, presented in this work, of the VBF-type $A$ pair production at a muon collider would be paramount to test the IDM as a possible underlying model of new physics. 
\edit{Nevertheless, our analysis can be impacted by systematic uncertainties. An accuracy below $5\%$ on the background is a necessity to achieve a sizeable significance in the direct detection channel. While such precision may be achievable for a muon collider in the future (especially given the clean environment), more detailed studies are required to improve our understanding of the background systematics. Our work offers motivation for targeting reduced uncertainties at the level of a few percent.}

Muon colliders have recently undergone a novel resurrection, with increased and renewed interest within the international collider community. The process we chose is typically suppressed at hadron as well as low-energy lepton colliders, and requires a high centre-of-mass energy collider with sufficiently high energies for the radiated gauge bosons. We demonstrated that a 10 \TeV machine is needed to discover or exclude the large number of benchmark points we proposed, rendering motivation for further investigation of the feasibility of such a machine.


\section*{Acknowledgements}
\sloppy{
We acknowledge support by the Deutsche Forschungsgemeinschaft (DFG, German Research Foundation) under Germany's Excellence Strategy --- EXC 2121 ``Quantum Universe'' --- 390833306. This work has been partially funded by the Deutsche Forschungsgemeinschaft (DFG, German Research Foundation) --- 491245950. J.B.\ is supported by the DFG Emmy Noether Grant No.\ BR 6995/1-1. TR is supported by the Croatian Science Foundation (HRZZ) under project HRZZ-IP-2022-10-2520. 
{J.B.\ thanks the Ruder Boskovic Institute for hospitality and support. TR thanks DESY Hamburg for their hospitality while parts of this work were completed.}
}


\appendix

\section{ML interpretation plots}
\label{sec:shap}

To understand the importance of each feature and identify the ones that contribute most to the decision of {\xgboost}, we turn to techniques used for explainable artificial intelligence. Shapley values~\cite{shapley}, which were initially introduced in the context of game theory, can be utilised to identify the gain from participating input variables in ML models and have been also explored within particle physics applications~\cite{Grojean:2020ech,Alasfar:2022vqw,Grojean:2022mef,Bahl:2023qwk}. For a model $f_F$ trained on a set of features $F$ and a feature of interest $m$, one can create subsets $S \subseteq F \setminus \{m\}$ that exclude $m$ and models $f_{S \cup \{m\}}$ and $f_S$ that are trained on the corresponding subsets. The effect of removing a feature is then obtained by the difference of the predictions and the Shapley value is the weighted average over all possible subsets
\begin{equation}
    \label{eq:shap}
    \phi_m = \sum_{S \subseteq F  \{m\}} \frac{\lvert S \rvert ! (\lvert F \rvert - \lvert S \rvert - 1)!}{\lvert F \rvert !} \left[ f_{S \cup \{m\}}(S \cup \{m\}) - f_S(S) \right]\;.
\end{equation}
The model yields large positive (negative) values when it predicts a signal (background) event. Therefore, a feature of interest $m$ that significantly impacts the predictions leads to Shapley values that deviate significantly from zero.

\begin{figure}[h!]
    \centering
    \includegraphics[scale=0.7]{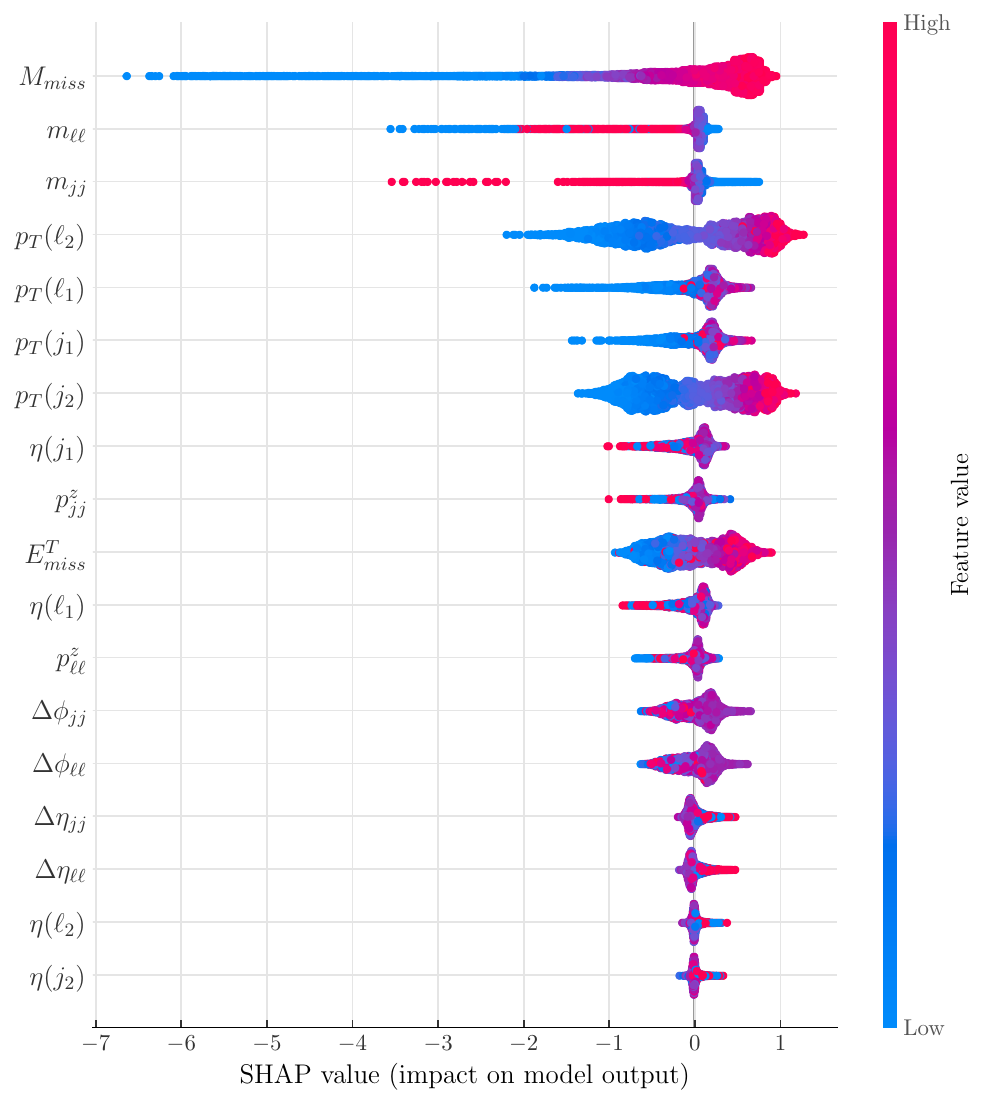}
    \caption{SHAP values for a collection of events evaluated for all the observables considered in our ML analysis. \label{fig:shap_beeswarm}}
\end{figure}

\begin{figure}
    \centering
    \includegraphics[width=0.48\textwidth]{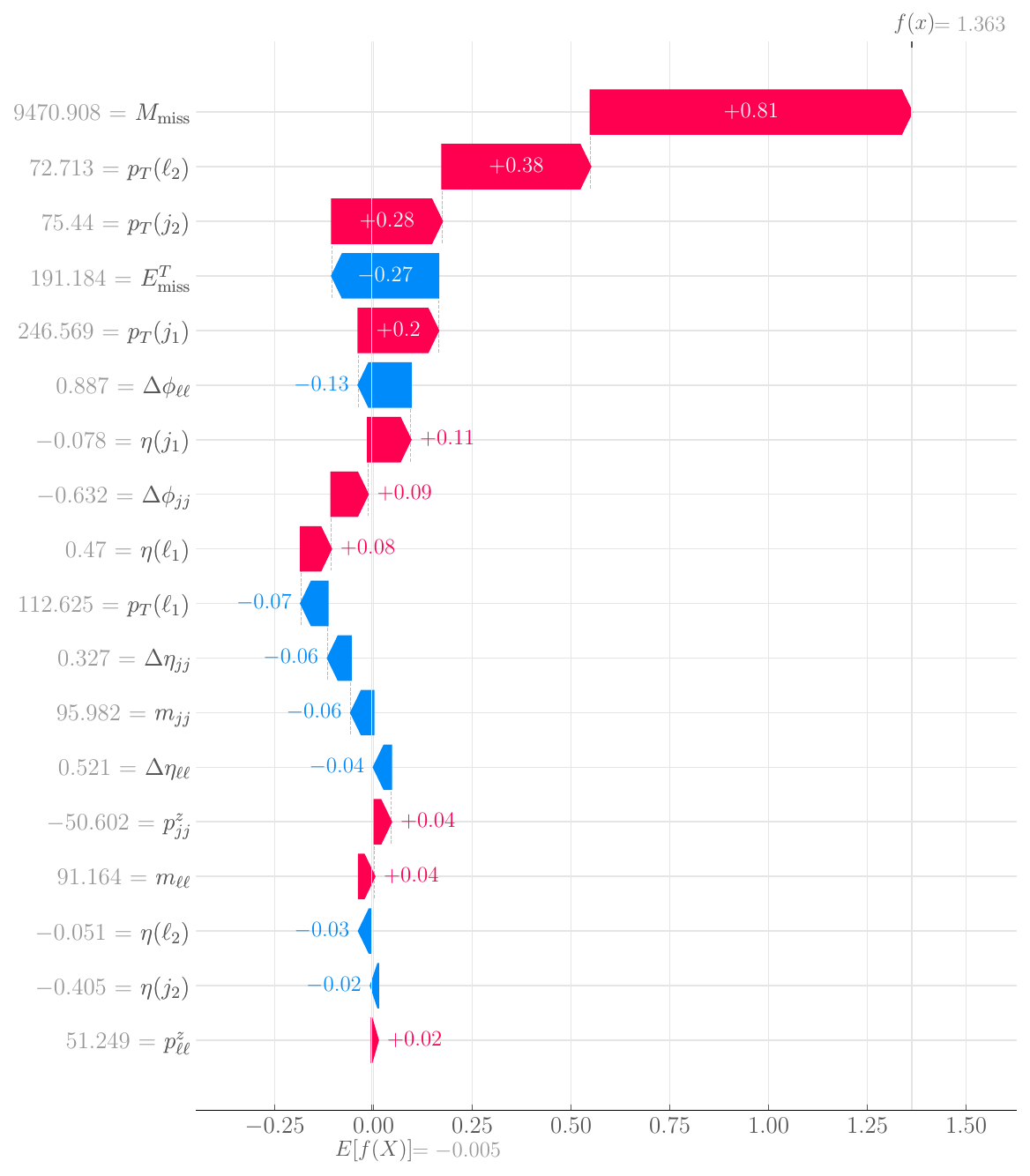}
    \includegraphics[width=0.48\textwidth]{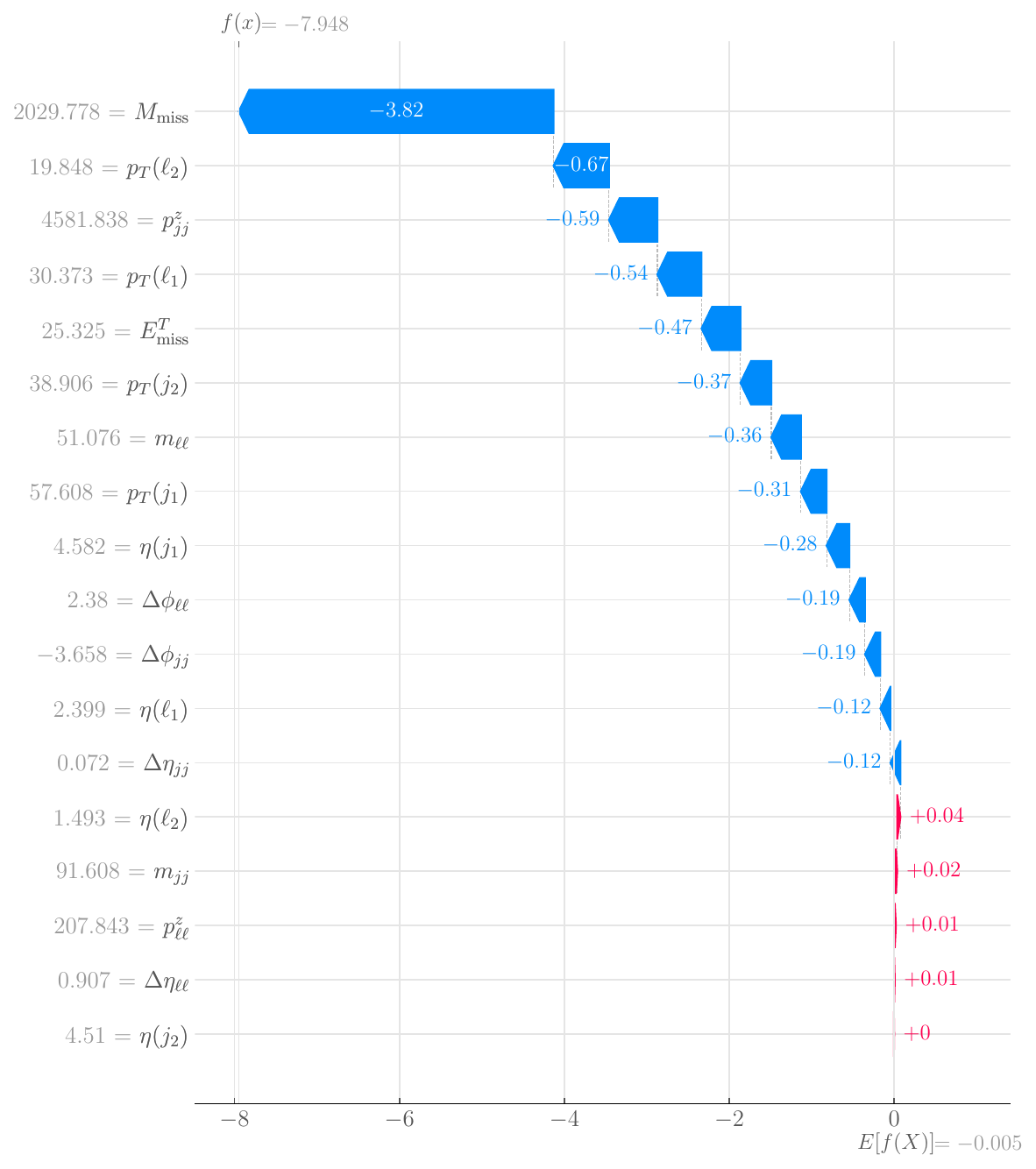}
    \caption{SHAP values for a signal (left) and a background (right) point.}
    \label{fig:shap_waterfall}
\end{figure}

While Shapley values provide a mathematically well-defined approach to fairly distribute importance attributions between input variables, they can be challenging to compute. We use the algorithms for explaining trees implemented in the {\textsc{SHAP}}~\cite{shap,shap_tree} package to obtain the importance of the input features for the test data of the BP1 point at $10$~TeV. The SHAP values are shown in \cref{fig:shap_beeswarm} for all the input features. It can be seen that large values of $M_\text{miss}$ are consistently classified as signal, while low values are decisively pushing the SHAP value to negative values which would imply that an event is characterised as background. For $m_{\ell\ell}$ the SHAP values indicate that values close to the $Z$ boson mass ($55 \lesssim m_{\ell\ell} \lesssim 95$ GeV) receive a positive attribution. This is relaxed for $m_{jj}$ where low values can still receive a positive contribution. These three variables along with the transverse momenta of the leptons and jets are particularly important for the classification of an event. It is indicated that the missing transverse energy is not as important as the aforementioned variables, which may seem surprising. However, it should be noted that $E^T_\text{miss}$ and $M_\text{miss}$ are correlated and a cut on the latter can already remove events that would be removed from $E^T_\text{miss}$ (this behaviour is also inferred from our cut-and-count analysis). The SHAP value for each observable for one signal and one background point is also shown as an example in \cref{fig:shap_waterfall}.

It should be noted that the usefulness of ML interpretations goes beyond the identification of important input features. Mistakes in the implementation that are normally hard to spot become transparent when the importance attribution of certain input variable is illogical. Furthermore, in cases with limited resources, unimportant features with almost no contribution to the ML output can be clearly identified and removed.   

\clearpage
\section{Dark matter properties of benchmark points}\label{sec:dmprops}

We here list the dark matter properties of the benchmark points, derived using \texttt{micrOMEGAs}. We note that all but BP1 and BP4 currently fulfil all dark matter constraints. In addition, as discussed in \cref{seq:procgen} a small decrease in $\lambda_{345}$ would also allow BP1 and BP4 to satisfy the latest direct detection bounds from the LUX-ZEPLIN experiment without significantly modifying the collider phenomenology of these points.

{\footnotesize
\begin{table}[h!]
\centering
\resizebox{\textwidth}{!}{%
\begin{tabular}{l c c c c c c c c}
\toprule
ID    &   $M_H$  &    $M_A$  & $M_{H^\pm}$  &  $\lambda_2$  & $\lambda_{345}$ & $\bar{\lambda}_{345}$ &  $\Omega_H\,h^2$ & $\sigma_{DD} [\pb]$ \\
\midrule
BP1   &  171.52  &  618.899  &  628.841     &  3.066190     &  0.14400 & 11.8307 & 0.000216& $6.25\times\,10^{-9}$ \\
BP2   & 766.72  &  964.775  &  974.106     &  1.495400     &  $-0.00590$ & 11.3277 & 0.000142&	$5.30\times\,10^{-13}$\\ 
BP3   &   60.975  &  496.049  &  498.244     &  2.337340     &  $-0.00480$ & 8.00454 &0.001263 &  $5.39\times\,10^{-11}$\\
BP4   &  59.000  &  300.700  &  316.100     &  0.188496     &  $-0.00384$ & 2.86943 &0.011338 &  $3.68\times\,10^{-11}$\\ 
BP5   &  60.905  &  400.325  &  406.473     &  3.430620     &  0.00396 & 5.17782 &0.001941 &  $3.68\times\,10^{-11}$\\
BP6   &   62.400  &  199.800  &  230.000     &  0.138230     &  0.00486 & 1.19550 &0.000223&   $5.28\times\,10^{-11}$\\ 
BP7   &  535.36  &  614.813  &  617.601     &  2.626370     &  $-0.00044$ & 3.01578 &0.000757&$6.03\times\,10^{-15}$\\
BP8   &  553.60  &  799.566  &  799.566     &  0.766550     &  $-0.01734$ & 10.9825 & 0.000120&	$8.77\times\,10^{-12}$\\
BP9   &  474.88  &  600.384  &  618.238     &  3.593980     &  $-0.00328$ & 4.45670 &0.000316 &	$4.26\times\,10^{-13}$  \\
BP10  &  501.76  &  670.165  &  678.137     &  2.827430     &  $-0.01498$ & 6.50753 & 0.000214&	$7.96\times\,10^{-12}$\\ 
BP11  &  736.00  &  941.656  &  947.464     &  0.942478     &  $-0.00926$ & 11.3933 &0.000138&	$1.42\times\,10^{-12}$\\
\bottomrule
\end{tabular}%
}
    \caption{Benchmark points used in this work. The last two columns show relic density and direct detection cross-sections, respectively.
   \label{tab:dmprops}}
\end{table}
}



\clearpage
\printbibliography[heading=bibintoc,title={References}]

\end{document}